\documentclass{article}

\PassOptionsToPackage{round,semicolon,authoryear}{natbib}

\usepackage[eandd, final]{neurips_2026}

\usepackage[utf8]{inputenc} 
\usepackage[T1]{fontenc}    
\usepackage[colorlinks=true, citecolor=blue, linkcolor=blue]{hyperref}

\newcommand{\subfigref}[2]{\hyperref[#1]{\figureautorefname~\ref*{#1}#2}}
\usepackage{booktabs}       
\usepackage{amsfonts}       
\usepackage{nicefrac}       
\usepackage{microtype}      
\usepackage{xcolor}         

\colorlet{revcolor}{black}   

\usepackage{amsmath}
\usepackage{graphicx}

\usepackage{float}

\title{\textsc{TurtLES}: A Large-Scale Benchmark for Turbulent 3D Neural PDE Surrogates}
  
\author{%
  Jarno Platenburg$^{1,2}$\thanks{Equal contribution. $^{\dagger}$Equal advising. Correspondence: \href{mailto:jarno.platenburg@gmail.com}{jarno.platenburg[at]gmail.com}, \href{mailto:armand.kassai@isir.upmc.fr}{armand.kassai[at]isir.upmc.fr}.} \quad Armand Kassa{\"\i} Koupa{\"\i}$^{1}$\footnotemark[1] \quad Paola Cinnella$^{3\dagger}$ \quad Patrick Gallinari$^{1,4\dagger}$ \\
  $^{1}$Sorbonne Universit\'e, CNRS, ISIR, Paris, France \\
  $^{2}$Delft University of Technology, Faculty of Aerospace Engineering, Delft, The Netherlands \\
  $^{3}$Sorbonne Universit\'e, CNRS, Institut Jean Le Rond d'Alembert, Paris, France \\
  $^{4}$Criteo AI Lab, Paris, France
}

\begin{document}

\maketitle

\begin{figure}[H]
    \centering
    \color{revcolor}
    \includegraphics[width=\linewidth]{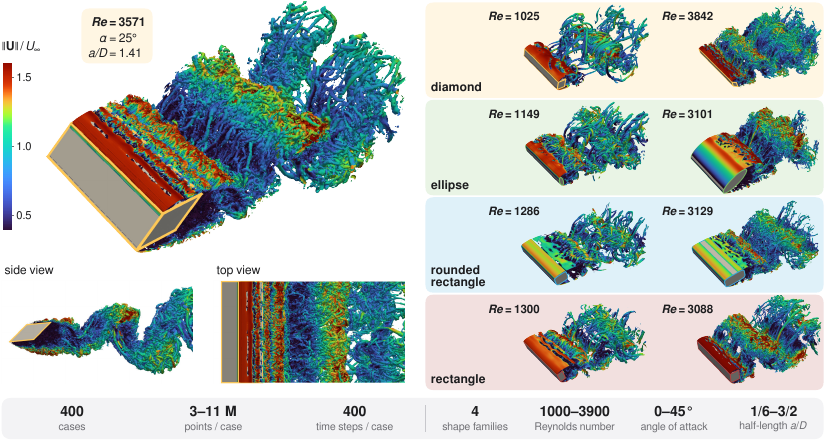} %
    \vspace{-.8em}
    \caption{\textsc{TurtLES} visualized through Q-criterion isosurfaces and its dataset parameters.}
    \label{fig:frontpage}
\end{figure}

\begin{abstract}
Machine-learning surrogates are increasingly used to accelerate computational fluid dynamics, yet progress is limited by the lack of benchmarks capturing realistic, time-dependent turbulent flows. We introduce \textcolor{revcolor}{\textsc{TurtLES}}, 	a \textcolor{revcolor}{14}~TB dataset of \textcolor{revcolor}{3D} turbulent cylinder wakes in the shear-layer transition regime\textcolor{revcolor}{, computed via high-fidelity implicit large-eddy simulations.} The dataset contains \textcolor{revcolor}{400} trajectories of 400 time steps each, with \textcolor{revcolor}{3--11} million points per frame, spanning variations in geometry, Reynolds number, and angle of attack. Unlike existing datasets, \textcolor{revcolor}{\textsc{TurtLES}} combines \textcolor{revcolor}{(i)~physically realistic turbulence with an active energy cascade driven by vortex stretching}, (ii)~dense irregular meshes, and (iii)~long-horizon temporal dynamics. We benchmark state-of-the-art neural operators on autoregressive rollouts and identify key failure modes, including temporal instability, loss of small-scale energy, and challenges in learning on large irregular domains. \textcolor{revcolor}{\textsc{TurtLES}} provides a new benchmark for evaluating surrogate models \textcolor{revcolor}{on 3D turbulent wakes at industrially relevant mesh resolution.} \textit{Project page}: \href{https://turtlesprojectml-alt.github.io/}{turtlesprojectml-alt.github.io}
\end{abstract}


\section{Introduction}
The simulation of physical phenomena is central to industrial workflows in design \citep{Allen2022}, forecasting \citep{Richardson1922WeatherPB}, control \citep{Brunton_Kutz_2022}, and decision-making \citep{GraphCast}. These phenomena are governed by partial differential equations describing the spatio-temporal evolution of physical fields. Because such systems are rarely analytically tractable, they are solved numerically, often at substantial computational cost \citep{ames2014numerical, Quarteroni2008NumericalAO}. Yet downstream decisions rarely require the full resolution produced by high-fidelity solvers \citep{airfrans2023, Zhiyin2015LES}. This mismatch has motivated surrogate models: low-cost approximations that reproduce task-relevant physical behavior while avoiding repeated full-scale simulation. Deep learning has emerged as a powerful framework for surrogate modeling \citep{Long2018, deBezenac2018}, with rapid progress in operator learning \citep{li2020fourier, Lu2019,serrano2024operator,Wu2024,Alkin2024}, neural simulation \citep{Brandstetter2022}, and generative emulation \citep{zebraserrano2024,Li2025DiT,enma}. Its adoption in realistic simulation settings, however, remains limited by a circular dependency: accurate surrogates require large, high-quality datasets, but such datasets must typically be generated by the same numerical solvers or physical experiments that surrogates are meant to replace.

Recent datasets have begun to address this bottleneck, but each leaves a critical axis uncovered. The Well~\citep{thewell2025} provides approximately 15~TB of dynamical trajectories across sixteen physical settings on regular grids, with most three-dimensional cases capped at $128^3$ resolution. DrivAerNet~\citep{drivaernet2025} reaches industrial-scale three-dimensional geometries with tens of millions of degrees of freedom on irregular meshes, but only at steady state. EAGLE~\citep{eagle2023} captures dynamic flows on irregular geometries, yet its $\mathcal{O}(10^3)$ points per frame are too coarse to resolve realistic three-dimensional flow structures. RealPDEBench~\citep{realpdebench2026} pairs real-world PIV measurements with matched simulations, but its fluid measurements are two-dimensional laser-sheet observations with only in-plane velocity components. It is therefore valuable for simulation--experiment alignment, but does not directly expose the three-dimensional velocity fields and multiscale structures needed to assess turbulent flow surrogates.

More broadly, many machine learning benchmarks conflate temporal complexity with turbulence. A flow may be unsteady, chaotic, or visually complex without containing the mechanisms that define three-dimensional turbulence. Several datasets described as ``dynamic fluid'' benchmarks fall into this category: they consist of unsteady laminar flows, chaotic two-dimensional systems, or three-dimensional dynamics dominated by buoyancy, radiative forcing, magnetic effects, or other mechanisms rather than by an inertial cascade \citep{pfaff2021traditionalcylinder, takamoto2024pdebenchextensivebenchmarkscientific}. In three dimensions, turbulence is sustained by vortex stretching, represented by the $(\boldsymbol{\omega}\cdot\nabla)\mathbf{u}$ term in the vorticity equation, which transfers kinetic energy from large eddies to progressively smaller scales until viscous dissipation closes the energy budget \citep{PopeBook}. In two dimensions, this mechanism is absent; in other forced systems, it may be secondary. A surrogate trained where vortex stretching is absent or marginal is therefore not learning three-dimensional turbulence, but rather interpolation of unsteady or chaotic fields without the cascade physics that define turbulent flow. Standard error metrics often hide this distinction, which becomes visible only through energy spectra, fluctuation statistics, and regime-transition behavior.

At the same time, the machine learning community has made substantial progress on the ingredients required for realistic surrogate modeling: long-horizon stability on regular grids \citep{mccabe2023multiple, lippe2023pde}, generalization \citep{huang2022metaautodecoder, Koupai2024GEPS, subramanian2023towards, takamoto2023learning}, neural operators for dense unstructured static fields \citep{li2023geometry, wen2025goat, luotransolverplus}, and generative emulators for chaotic dynamics \citep{zhou2025reframinggenerativemodelsphysical, pedersen2025thermalizer}. What remains missing is a benchmark that stresses these capabilities jointly: dynamic three-dimensional trajectories, dense irregular discretizations, long temporal horizons, and physically turbulent flow driven by vortex stretching.

To fill this gap, we introduce \textsc{TurtLES}, a \textcolor{revcolor}{14}~TB dataset of high-fidelity simulations of three-dimensional cylinder wakes in the shear-layer transition regime. The dataset contains \textcolor{revcolor}{400} trajectories of 400 time steps each, with \textcolor{revcolor}{3--11} million points per frame sampled on irregular point clouds. Across trajectories, we vary the cylinder cross-section, Reynolds number, and angle of attack, producing geometry-dependent wakes that span subcritical to transitional regimes. In this range, shear-layer transition generates a turbulent near wake, so the dataset contains an active three-dimensional cascade in the simulated fields themselves.

\textsc{TurtLES} makes three contributions to surrogate modeling for PDEs. First, it provides dynamic three-dimensional flow data on dense irregular discretizations, reaching millions of points per frame rather than relying on regular grids or coarse point clouds. Second, it enables long-horizon evaluation over hundreds of autoregressive time steps in chaotic, geometry-dependent wakes. Third, it supports physically grounded assessment of turbulent surrogates through high-frequency quantities such as fluctuation statistics, near-wake spectra, and intermittent shedding-mode behavior. By combining \textcolor{revcolor}{industrially relevant mesh resolution}, long temporal trajectories, irregular geometry, and physically turbulent flow, \textsc{TurtLES} targets a regime that existing benchmarks do not jointly cover. It exposes whether learned surrogates can remain stable, resolve multiscale turbulent content, and generalize across geometry-dependent flow regimes, rather than only minimizing standard field-wise errors.


\section{Related Work}
\paragraph{Operator Learning}
Operator learning provides a way to train neural networks to approximate PDE solution maps directly between function spaces. Early neural operators such as DeepONet and FNO showed that these models can act as mesh-independent surrogates for PDEs \citep{Lu2019,li2020fourier}. Since then, the field has moved toward models that are more accurate on complex geometries, better at resolving multiscale and high-frequency features, and scalable to very large meshes.

Recent work has explored several directions. Multiresolution and wavelet-based operators improve the representation of fine-scale structures \citep{gupta2021multiwavelet}, while latent and geometry-aware formulations extend neural operators to irregular domains \citep{Tran2023ffno,li2023geometry}. Attention-based models such as OFormer, GNOT and Transolver further improve operator learning on point clouds and arbitrary domains by using graph or transformer encoders to handle irregular inputs and complex PDE dynamics \citep{li2023transformer,hao2023gnot,Wu2024}. However, full attention remains expensive for high-resolution fields.

This has motivated more efficient neural operator surrogates. Perceiver-style and latent-token models such as Aroma and UPT compress large inputs into compact latent representations, reducing the cost of learning on dense meshes \citep{aroma2024,Alkin2024}. \citet{calmpde, koupai2025efficientgenerativetransformeroperators} proposed continuous convolution for learning hierarchical neural operators that can operate on very dense grids, allowing million point trajectory generation. \citet{wen2025goat} uses geometry-aware multiscale graph/attention encoding to scale neural PDE surrogates to arbitrary large meshes, including industrial 3D computational fluid dynamics (CFD) domains.

\paragraph{Datasets for Machine Learning Applications}
A key bottleneck in training neural operators for fluid dynamics is the limited availability of high-quality datasets. Generating such data requires solving the Navier-Stokes equations at high spatial and temporal resolution, which comes with substantial computational and storage costs. As a result, most existing datasets are restricted in either fidelity, dimensionality, or physical complexity.

A substantial body of work \citep{airfrans2023, BlendedNet2025, drivaernet2025, unifoil2025, emmiwing2025} has focused on performing steady-state flow simulations through the Reynolds-averaged Navier-Stokes (RANS) equations, where all turbulence fluctuations are modeled rather than computed directly. While these approaches significantly reduce computational complexity, RANS-based simulations suffer from well-documented limitations in capturing complex phenomena such as sharp pressure gradients, laminar-to-turbulent transition and flow separation \citep{wilcox1998turbulence}. Furthermore, by definition, steady-state solutions neglect temporal dynamics, preventing operators from learning transient flow phenomena.

Therefore, several benchmarks based on unsteady flows have been introduced. A common cost-reducing strategy is to restrict simulations to two-dimensional domains \citep{huang2020, Viquerat2020, Jakob2020, pfaff2021traditionalcylinder, han2022predictingphysicsmeshreducedspace, megaflow2d2023, eagle2023, flowbench2024, cfdbench2024, FluidGPT2025}, at the expense of losing relevant three-dimensional flow phenomena. Alternatively, higher-fidelity transient datasets generated using large-eddy simulation (LES) or direct numerical simulation (DNS) have been explored in well-known (multi-)physics benchmarks \citep{bubbleml2023, blastnet202023, apebench2024, takamoto2024pdebenchextensivebenchmarkscientific, thewell2025, realm2026}. There, they are typically limited to regular grids and idealized configurations (e.g., homogeneous isotropic turbulence, no objects within the domain, periodic boundary conditions). These flow problems do not introduce geometric complexities or capture the anisotropic nature of real-world turbulent flows. It should be noted that scale-resolving simulations for machine learning applications have been conducted by \citet{ahmedml2024, windsorml2025} on the Ahmed and Windsor car bodies, respectively. However, due to storage constraints, only time-averaged volume data was made publicly available.

To the best of our knowledge, there is no dataset that combines three-dimensional, time-resolved high-fidelity CFD simulations with systematic variations in complex geometries while fully resolving a broad range of turbulent scales. The present study aims to address this gap by performing implicit large eddy simulations on irregular grids for the canonical cylinder flow case across a range of inflow and geometric configurations.


\section{Dataset Presentation}
The dataset captures a wide range of fluid--object interactions and wake dynamics through systematic variations in geometric parameters and inflow conditions\textcolor{revcolor}{, as illustrated in \autoref{fig:frontpage} and \autoref{app:datavis}}. To accurately resolve the resulting flow features and their temporal evolution, costly high-fidelity numerical simulations are performed. Generating a large number of such trajectories requires a robust and fully automated data generation pipeline. The remainder of this section describes the components of this pipeline, including the governing fluid flow model, the definition of the parameter sampling space, the automated simulation setup, and the subsequent data processing.

\paragraph{Fluid flow simulation}
The fluid dynamics are governed by the Navier--Stokes equations, a system of equations that expresses the conservation of mass, momentum, and energy for a compressible fluid. Given a set of inflow and boundary conditions, characterized by the Reynolds number $Re$, the Mach number $Ma$ and Prandtl number $Pr$, these equations describe how a fluid evolves over space and time. In this study, we focus our attention on the evolution of the pressure $p$, density $\rho$ and velocity components $u$, $v$, $w$. 

At sufficiently high Reynolds numbers, the Navier-Stokes equations give rise to turbulence, whose multi-scale, chaotic dynamics make direct numerical solution computationally prohibitive. We solve these equations using an implicit Large Eddy Simulation (iLES) approach, in which the most energetic turbulent scales are computed directly, while the truncation error of the numerical scheme acts as a model for the behavior of the smaller, unresolved scales. In this context, the Flux Reconstruction (FR) scheme \citep{Huynh2007} has proven to be effective and is implemented using PyFR \citep{WITHERDEN20143028}, an open-source, high-order accurate GPU-based solver. The reader is referred to \autoref{sec:appquan} for quantity definitions, and to \autoref{sec:appiLES} for a detailed description of the compressible Navier-Stokes equations and the solution method used in this study. 

\begin{figure}[t]
    \centering
    \includegraphics[width=\linewidth]{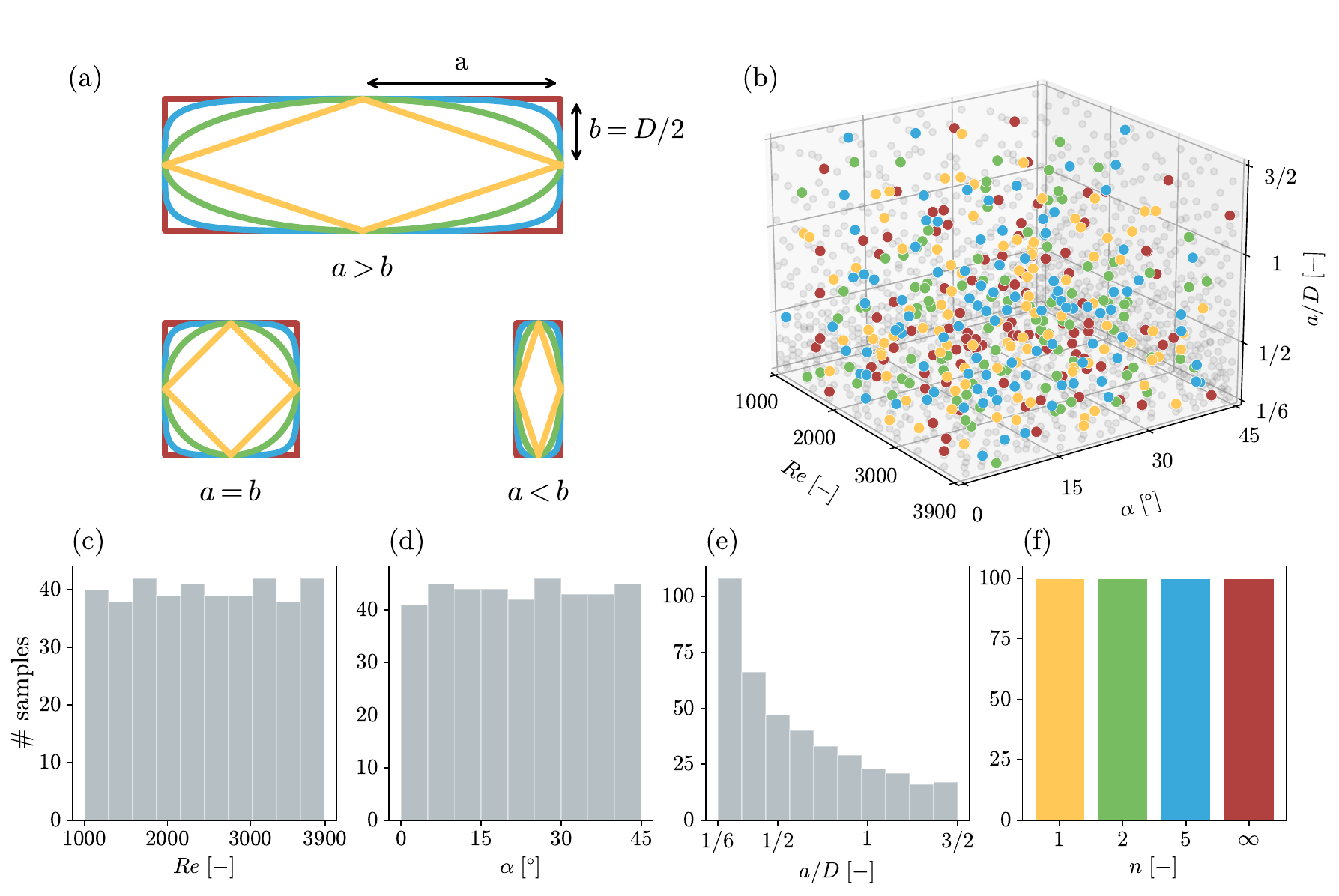} %
    \caption{Overview of the parameter sampling space. (a) Definition of the super-elliptical geometry. (b) The \textcolor{revcolor}{$N_s=400$} samples visualized in three dimensions, including projections on the $x,y,z$ planes in gray for reference. (c-f) Histograms of the sampled parameters. Note that the colors used for different values of $n$ in (f) correspond with those in (a) and (b).}
    \label{fig:parameterspace}
\end{figure}

\paragraph{Sampling space} \label{sec:appsample}
The dataset is constructed within a four-dimensional parameter space. Variations in geometry are defined by the super-ellipse (or Lamé curve), \textcolor{revcolor}{i.e.}\ all points $(x,y)$ satisfying: 
\begin{equation} \label{eq:lame}
    \left| \frac{x}{a}\right|^n + \left| \frac{y}{b}\right|^n = 1,
\end{equation}
where $a$ and $b$ represent half the horizontal and vertical dimensions of the object, respectively, as illustrated in \subfigref{fig:parameterspace}{a}. The aspect ratio is defined as $AR = a/b$. To reduce the number of variable parameters, $b=1/2$ is fixed for all cases\textcolor{revcolor}{, giving a constant object thickness $D = 2b = 1$.} The semi-axis $a$ is sampled log-uniformly over the range $[1/6, 3/2]$, which results in aspect ratios spanning from 1/3 to 3, with an equal proportion of cases having $AR<1$ and $AR>1$. The exponent $n$ is selected from the discrete set $\{1, 2, 5, \infty\}$, corresponding to diamond, elliptical, rounded rectangle and rectangular shapes, respectively. 

The inflow conditions are characterized by Reynolds numbers $Re \in [1000, 3900]$, where the length scale used in the definition of $Re$ is \textcolor{revcolor}{the object thickness $D$}. Within this range, the wake \textcolor{revcolor}{transitions to turbulence} while the boundary layer remains laminar, consistent with observations from circular cylinder flow studies \citep{Lienhard1966}. The inflow angle of attack $\alpha$ is the counter-clockwise angle between the $x$-axis and the object’s centerline. It is varied between $0$ and $45$ degrees, introducing unique flow separation behavior for each configuration. Limiting the range to $45^\circ$ prevents generating geometrically redundant cases. Both $Re$ and $\alpha$ are varied uniformly and restricted to integer values for simplicity.

To obtain representative samples of the parameter space, we employ a Sobol' sequence \citep{SOBOL196786}, which is a quasi-random low-discrepancy sequence designed to minimize clustering and gaps in high-dimensional sampling. Unlike related methods such as Latin Hypercube Sampling, it supports sequential sampling, making it well suited for dataset generation where the total number of cases need not be specified in advance. A four-dimensional unit hypercube is sampled \textcolor{revcolor}{$N_s = 400$} times and subsequently transformed to the previously defined (log-)uniform or discrete parameter ranges. The resulting space is visualized in \subfigref{fig:parameterspace}{b} together with histograms of the different parameters in \subfigref{fig:parameterspace}{c--f}. \textcolor{revcolor}{Trajectories are partitioned into fixed training, validation and test sets of 320/40/40 cases. The split is released with the dataset and described in \autoref{sec:appsplit}.}

\begin{figure}[t]
    \centering
    \begin{minipage}[b]{0.45\textwidth}
        \centering
        \includegraphics[width=\textwidth]{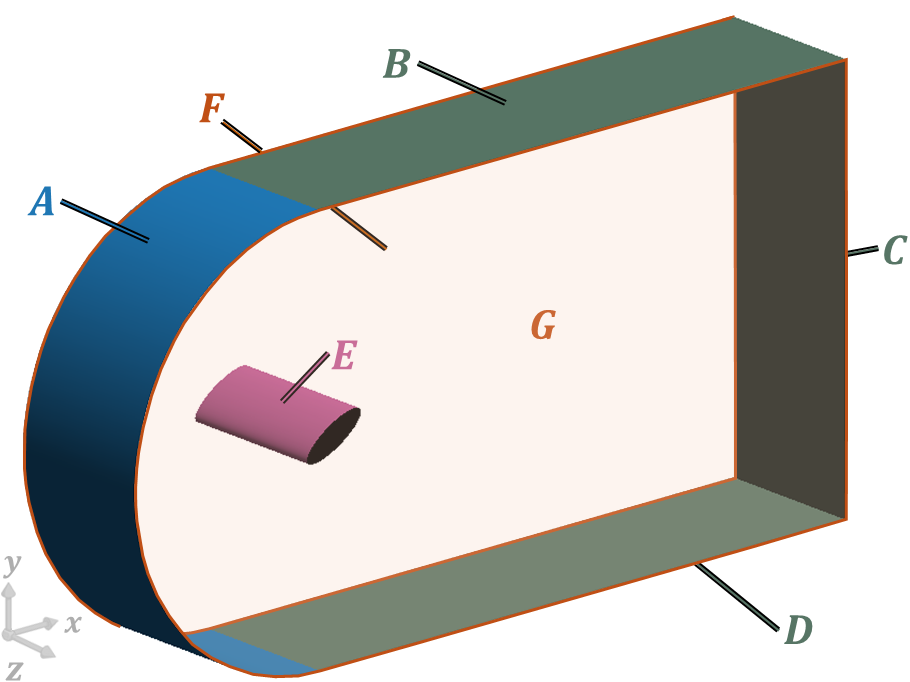}
        \\ (a) 
    \end{minipage}
    \hfill
    \begin{minipage}[b]{0.52\textwidth}
        \centering
        \includegraphics[width=\textwidth]{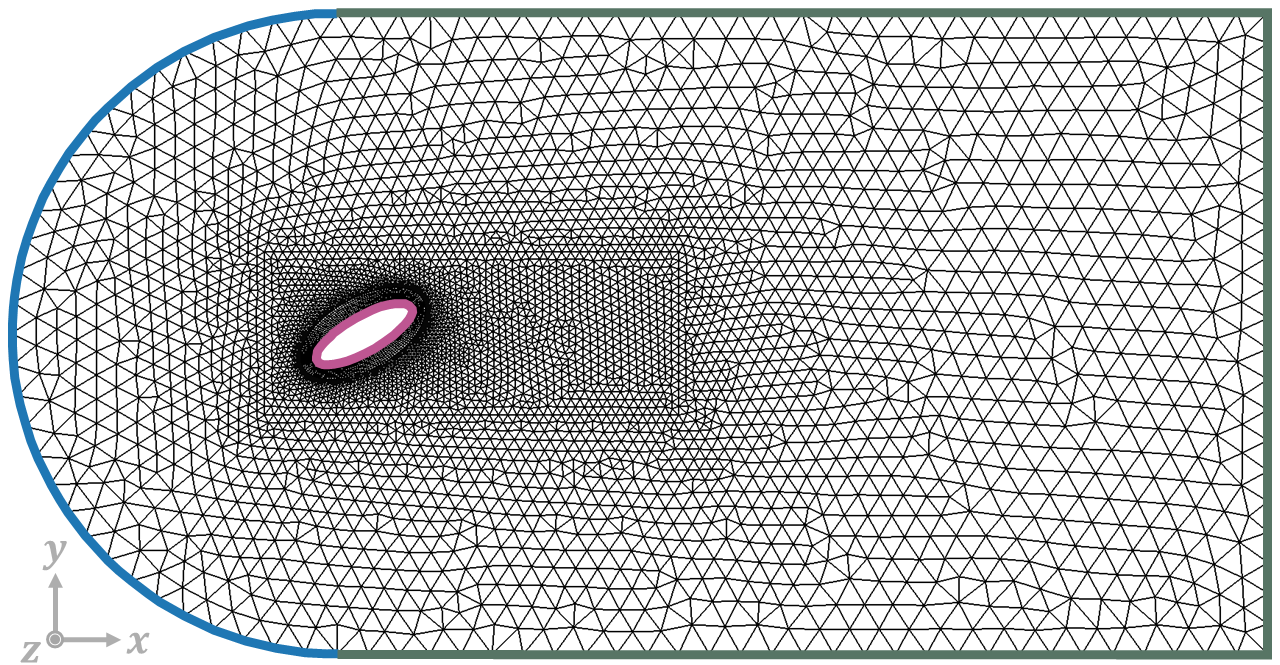}
        \\ (b) 
    \end{minipage}
    \caption{Schematic overview of the domain. (a) Three-dimensional view illustrating the physical boundaries, including the inlet (\textbf{A}), outlets (\textbf{B}--\textbf{D}), object surface (\textbf{E}), and periodic side-walls (\textbf{F}, \textbf{G}). (b) $xy$-plane of the computational mesh, which is extended into three dimensions via a structured extrusion in the span-wise ($z$) direction.
    }
    \label{fig:mainDomains}
\end{figure}

\paragraph{Simulation setup}
For each case, the super-elliptical object is placed inside the 3D computational domain in \subfigref{fig:mainDomains}{a} at an angle $\alpha$. The flow is prescribed through boundary conditions placed on the walls (\textbf{A}--\textbf{G}). \textbf{A} is the inlet, \textbf{B}--\textbf{D} are the outlet and \textbf{E} is the object surface. Walls \textbf{F} and \textbf{G} represent the side-walls of the domain. They are assigned periodic boundary conditions, which means flow leaving \textbf{F} enters back through \textbf{G}. An automated routine is implemented to generate the computational mesh using Gmsh \citep{GMSH}. In \subfigref{fig:mainDomains}{b}, we show the resulting mesh in the xy-plane for the domain presented in \subfigref{fig:mainDomains}{a}. This two-dimensional mesh is then extended into three dimensions by a 30-layer structured extrusion in the span-wise direction. The flux reconstruction method places solution nodes inside these elements. A single element contains 75-125 points, yielding \textcolor{revcolor}{2.9M to 10.5M unique solution points per case within the stored region (see \textit{Data processing} below)}. Additional information on the domain definition and meshing procedure are given in \autoref{sec:appsimset}. 

The numerical framework is validated against four benchmark cases: a circular cylinder ($a = 0.5$, $n = 2$) at $Re  = 1000$ and $Re = 3900$, a square cylinder ($a = 0.5$, $n = \infty$) at $Re = 2000$ and an elongated rectangle ($a = 2.5$, $n = \infty$) at $Re = 3000$. The results indicate good agreement with DNS and experimental data reported in the literature. A detailed comparison is provided in \autoref{sec:appvali}.

\begin{figure}[t]
    \centering
    \includegraphics[width=\linewidth]{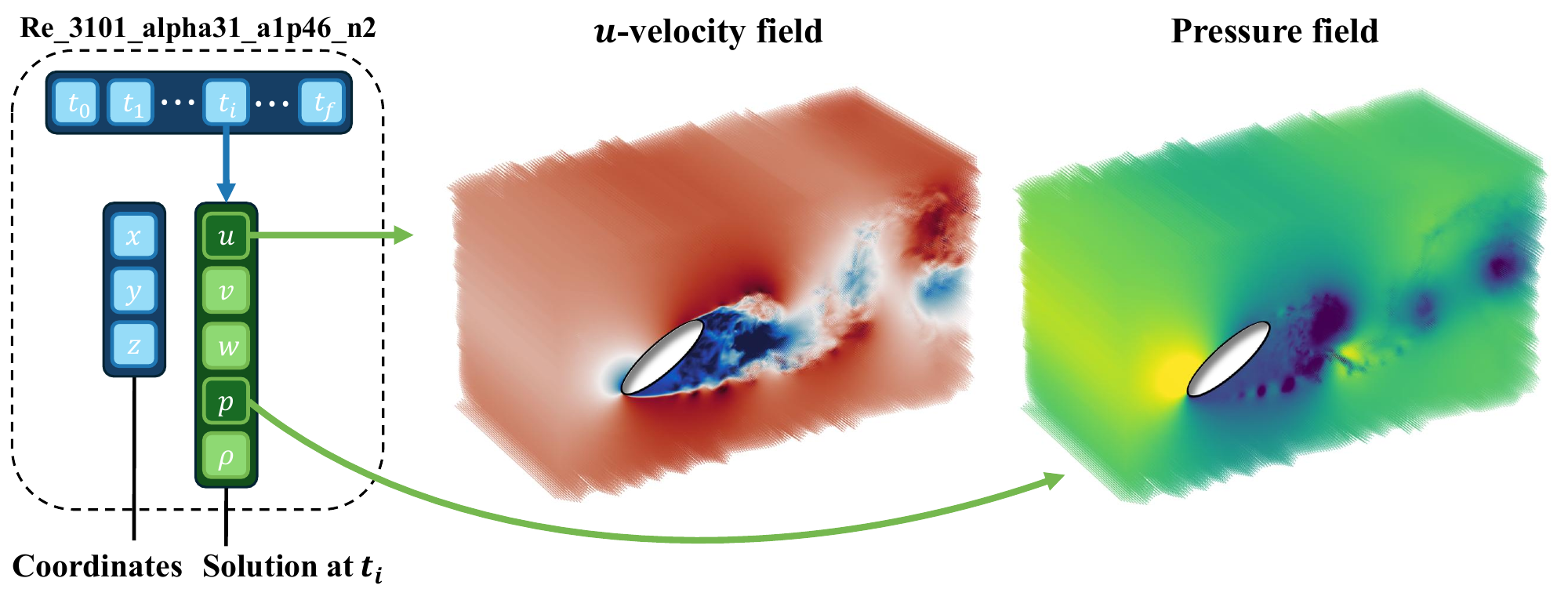}
    \caption{Overview of the flow case data structure. For each case, a single set of coordinates is stored with the solutions at 400 time-steps (e.g. $t_0, t_1, ..., t_f$, where $t_f = 399$). A solution comprises the three-dimensional pressure ($p$), density ($\rho$), and velocity ($u$, $v$, $w$) fields.}
    \label{fig:datastruc}
\end{figure}

\paragraph{Data processing}
Processing steps are applied to limit the total data size while ensuring the solutions remain physically representative. Data collection starts only after the initial transient has been washed away according to the numerical convergence statistic proposed by \citet{Boudet2015}. Subsequently, 400 three-dimensional snapshots are saved of the three velocity components, pressure and density: $\{u, v, w, p, \rho\}$ at fixed time intervals $\Delta t_s$. A snapshot is restricted to all solution nodes inside the rectangular box given by $x/D \in [-3, 10]$, $y/D \in [-3.5, 3.5]$ and $z/D \in [0, 2\pi]$. The sampling interval $\Delta t_s$ exceeds the solver time-step to manage storage constraints but is chosen such that the inertial range of the turbulence spectrum is retained for all cases. Ultimately, one trajectory covers a time-period in which most characteristic frequencies are repeated multiple times. For instance, four vortex shedding cycles are captured during $400 \Delta t_s$ in the \textcolor{revcolor}{circular} cylinder case ($a = 0.5, n = 2$). \autoref{fig:datastruc} visualizes the data structure of a single flow case. A case directory contains one set of node coordinates, as the grid remains constant throughout the entire trajectory, accompanied with the corresponding solutions for all 400 time-steps. \textcolor{revcolor}{Each case directory additionally stores power spectra of the streamwise ($u$) and cross-stream ($v$) velocity components at three centerline ($y/D=0$) probes located behind each object, together with the script used to compute them, providing per-case evidence of the turbulent cascade and a physically grounded target for surrogate evaluation.} \autoref{sec:appproc} goes into more detail on the statistic used to measure the initial transient, the definition of $\Delta t_s$ and \textcolor{revcolor}{the computation of the velocity power spectra.}

\textcolor{revcolor}{To lower the barrier to entry, we additionally define \textsc{mini-TurtLES}: a set of standardized subsets that reduce the dataset in tiers down to 0.8~TB, together with a 5~GB debug subset. Subset definitions and access details are given in \autoref{sec:appmini}.}

\section{Benchmarking Setup}
\paragraph{Motivation and benchmark.}
To better demonstrate the value of this dataset, we evaluate the performance of current neural operators on these data. The machine learning task considered here is temporal prediction of the \textcolor{revcolor}{fields $p$, $\rho$, $u$, $v$, and $w$. The dataset is designed to expose limitations of existing neural operators for dynamic forecasting}: handling dense meshes remains computationally expensive, stability over long rollout horizons is often poor, and high-frequency features are frequently underrepresented or smoothed out. Since the dataset is parametric, it also tests generalization across a rich distribution of physical configurations.

We evaluate several recent state-of-the-art neural operator architectures from the literature: UPT \citep{Alkin2024}, GINO \citep{li2023geometry}, Transolver++ \citep{luotransolverplus}, and the continuous layer proposed in CALM-PDE \citep{calmpde}, combined with the scaling improvements introduced in ECHO for larger meshes \citep{koupai2025efficientgenerativetransformeroperators}. \textcolor{revcolor}{These baselines are intended as an initial benchmark, illustrating the level of performance attainable with well-established models under a shared, fixed training budget. Architectures, hyperparameters and model-specific adjustments are documented in \autoref{app:baselines}.}

\paragraph{Training setup.}
The models are trained \textcolor{revcolor}{on all 320 training trajectories (\autoref{sec:appsplit})} to predict the temporal residual between time \(t\) and time \(t+1\) from a single input frame:
\[
    \mathrm{NO}(u_t, Re) + u_t \mapsto u_{t+1},
\]
where the Reynolds number \(Re\) is also provided as an input. Learning the residual is generally easier for neural networks than directly predicting the full next state. The training loss is a standard mean squared error, computed on the predicted residual. Since the meshes are very large, a subset of points is sampled during training, and each model predicts the corresponding points at the next time step. All models are trained for \(100{,}000\) steps with a batch size of \(2\), for approximately \(20\) hours on a single H100 GPU. We use the AdamW optimizer with a weight decay of \(10^{-3}\) and a learning rate of \(10^{-4}\). All models are trained with a linear warmup over the first \(5{,}000\) steps, followed by a cosine learning-rate scheduler.
\paragraph{Inference and evaluation setup.}
At inference time, each model rolls out the full dynamics using only the initial condition and the Reynolds number. Each trajectory contains \(400\) time steps, while the number of mesh nodes varies across trajectories; we therefore use a batch size of \(1\) and predict all nodes in each frame by iterating over subsets of points, since the full mesh cannot be processed in a single forward pass. The rollout produces predictions \(\hat{u}_{t,n,c}\) and reference values \(u_{t,n,c}\), where \(t \in \{1,\dots,T_{\mathrm{out}}\}\) indexes rollout steps, \(n \in \{1,\dots,N\}\) indexes mesh nodes, and \(c \in \{1,\dots,C\}\) indexes physical channels, with \(C=5\) corresponding to \(p,\rho,u,v,w\). 

\textcolor{revcolor}{We report four quantities computed from the autoregressive rollout. The first two are the relative \(L^2\) error and the variance-scaled RMSE (VRMSE). We use VRMSE because the physical channels span several orders of magnitude, so unnormalized RMSE would be dominated by \(p\) and \(\rho\):}
\[\mathrm{Rel}\text{-}L^2=\frac{1}{T_{\mathrm{out}}}\sum_{t=1}^{T_{\mathrm{out}}}\frac{\left\lVert \hat{u}_t-u_t \right\rVert_2}{\left\lVert u_t \right\rVert_2+\varepsilon}, \qquad \mathrm{VRMSE} = \frac{1}{T_{\mathrm{out}}} \sum_{t=1}^{T_{\mathrm{out}}} \frac{\left\lVert \hat{u}_t-u_t \right\rVert_2}{\left\lVert u_t-\bar{u}_t \right\rVert_2+\varepsilon}, \qquad \varepsilon=10^{-8},\]
where \(\bar{u}_t\) denotes the per-channel spatial mean of \(u_t\); a predictor that outputs this per-frame spatial mean attains \(\mathrm{VRMSE}=1\). 

\textcolor{revcolor}{Third, to assess multi-scale accuracy, we compare spanwise energy spectra of the streamwise velocity $u$. The computational mesh has 121 uniformly spaced spanwise solution points, periodic in $z \in [0, 2\pi]$, and thus requires no interpolation. For each frame, we take the orthonormalized real FFT of $u$ along $z$ and average the squared modulus over all $(x,y)$ locations to obtain the power spectrum}
\[\textcolor{revcolor}{P_t(k_z) = \frac{1}{N_{xy}} \sum_{(x,y)} \left|\hat{u}_{t}(x,y,k_z)\right|^2.}\]
\textcolor{revcolor}{We split the wavenumber range in $\log k_z$ into low-, mid-, and high-frequency bands, $B_{\mathrm{low}} = \{1,2,3\}$, $B_{\mathrm{mid}} = \{4,\dots,15\}$, $B_{\mathrm{high}} = \{16,\dots,60\}$, and report the RMS log-power discrepancy per band:}
\[
\textcolor{revcolor}{\mathrm{specRMSE}_{B} = \frac{1}{T_{\mathrm{out}}}\sum_{t=1}^{T_{\mathrm{out}}}\sqrt{\frac{1}{|B|}\sum_{k_z\in B}\left(\log P^{\hat{u}}_{t}(k_z)-\log P^{u}_{t}(k_z)\right)^2}.}
\]
These three bands probe large, intermediate, and small flow scales, respectively, and penalize spectral mismatches that pointwise RMSE can average away.

Finally, we report \textcolor{revcolor}{inference cost as the GPU time required to predict one trajectory on the full mesh.} Encoder--decoder baselines such as UPT and CALM-PDE encode the input mesh once and decode query points in chunks, whereas Transolver\textcolor{revcolor}{++} processes the mesh through a tile partition; all models use the same per-frame timing protocol, differing only in kernel-launch granularity. 


\section{Results}
\label{sec:results}

We evaluate the models along three complementary axes: rollout accuracy, spectral fidelity, and latency. These criteria are designed to separate errors that are often conflated in field-wise metrics. A model may achieve low pointwise error while dissipating small-scale turbulent content, or may perform well over short horizons while drifting during long autoregressive rollouts.

\paragraph{Dynamics forecasting.}
In \autoref{tab:main_results}, we report the models' performance on the first frame prediction. \textcolor{revcolor}{Under the shared training budget, CALM-PDE obtains the best field-level accuracy among the four baselines}, with the lowest relative $L^2$ error ($6.80\times 10^{-2}$) and the lowest VRMSE ($7.26\times 10^{-2}$). UPT is close behind in relative $L^2$ error ($7.01\times 10^{-2}$), but its VRMSE is about 23\% higher ($8.94\times 10^{-2}$), indicating larger errors once the spatial variance of each frame is accounted for. Transolver++ and GINO are substantially less accurate in this setting: both have higher relative $L^2$ errors, and their VRMSE values are at or above the mean-field level ($1.06$ and $1.21$, respectively), i.e.\ no better than predicting the per-frame spatial mean. The spectral metrics further show that UPT and CALM-PDE preserve the low-, mid-, and high-frequency content with similar accuracy, while Transolver++ exhibits a much larger high-frequency error and GINO a larger low-frequency error.

\begin{table}[t]
    \centering
    \caption{Next frame results on the test set. Lower is
    better for all metrics. Spectral errors are RMS log-power discrepancies computed over low-, mid-, and high-frequency \textcolor{revcolor}{spanwise} wavenumber bands.}
    \label{tab:main_results}
    \small
    \setlength{\tabcolsep}{5.5pt}
    \begin{tabular*}{\linewidth}{@{\extracolsep{\fill}}lccccc@{}}
        \toprule
        & \multicolumn{2}{c}{Field error}
        & \multicolumn{3}{c}{Spectral error} \\
        \cmidrule(lr){2-3}
        \cmidrule(l){4-6}
        Model
        & Rel-$L^2$ $\downarrow$
        & VRMSE $\downarrow$
        & Low $\downarrow$
        & Mid $\downarrow$
        & High $\downarrow$ \\
        \midrule
        UPT          & 7.01e-2 & 8.94e-2 & 4.69e-2 & 2.33e-2 & 4.42e-2 \\
        GINO         & 1.32e-1 & 1.21    & 1.32e-1 & 7.31e-2 & 4.48e-2     \\
        Transolver++ & 1.08e-1 & 1.06    & 8.02e-2 & 2.89e-2 & 4.41e-1 \\
        CALM-PDE     & 6.80e-2 & 7.26e-2 & 4.66e-2 & 2.38e-2 & 4.44e-2      \\
        \bottomrule
    \end{tabular*}
\end{table}

\paragraph{Long-range stability.}
\autoref{tab:temporal_stability} highlights a clear difference between short-horizon accuracy and long-horizon rollout stability. CALM-PDE remains stable throughout the autoregressive rollout, with VRMSE increasing only from $4.61\times 10^{-1}$ in the first $50$ steps to $7.89\times 10^{-1}$ in the last rollout window. In contrast, UPT begins with a moderate error but drifts rapidly over time, reaching a VRMSE of $5.12$ in the final window. Transolver++ is unstable from the beginning of the rollout and its error grows sharply, reaching $31.24$ over steps $301$--$399$. These results indicate that CALM-PDE better preserves the dynamics of the wake attractor over long autoregressive horizons, whereas the other reported baselines accumulate errors that eventually dominate the trajectory. 

\begin{table}[t]
    \centering
    \caption{Rollout stability and computational cost. Rollout columns: VRMSE averaged over windows of rollout steps. \textcolor{revcolor}{Cost: GPU time to generate (solver) or predict (surrogates) one full trajectory, i.e.\ wall-clock time $\times$ GPUs used (four for the solver, one for each surrogate) on NVIDIA H100 80~GB.}}
    \label{tab:temporal_stability}
    \small
    \setlength{\tabcolsep}{5pt}
    \begin{tabular*}{\linewidth}{@{\extracolsep{\fill}}lccccc@{}}
        \toprule
        & \multicolumn{4}{c}{Rollout VRMSE $\downarrow$}
        & \\
        \cmidrule(lr){2-5}
        Model
        & $1$--$50$
        & $51$--$150$
        & $151$--$300$
        & $301$--$399$
        & \textcolor{revcolor}{Cost [GPU-s] $\downarrow$} \\
        \midrule
        \textcolor{revcolor}{Solver (PyFR)} & -- & -- & -- & -- & \textcolor{revcolor}{13160} \\
        \midrule
        UPT          & 6.51e-1 & 1.56    & 3.31    & 5.12    & \textcolor{revcolor}{835}  \\
        GINO         & 3.18    & 10.32   & 24.31   & 33.12   & \textcolor{revcolor}{4097} \\
        Transolver++ & 2.33    & 8.32    & 19.33   & 31.24   & \textcolor{revcolor}{1148} \\
        CALM-PDE     & 4.61e-1 & 5.46e-1 & 6.52e-1 & 7.89e-1 & \textcolor{revcolor}{1073} \\
        \bottomrule
    \end{tabular*}
\end{table}

\begin{figure}[t]
    \centering
    \includegraphics[width=\linewidth]{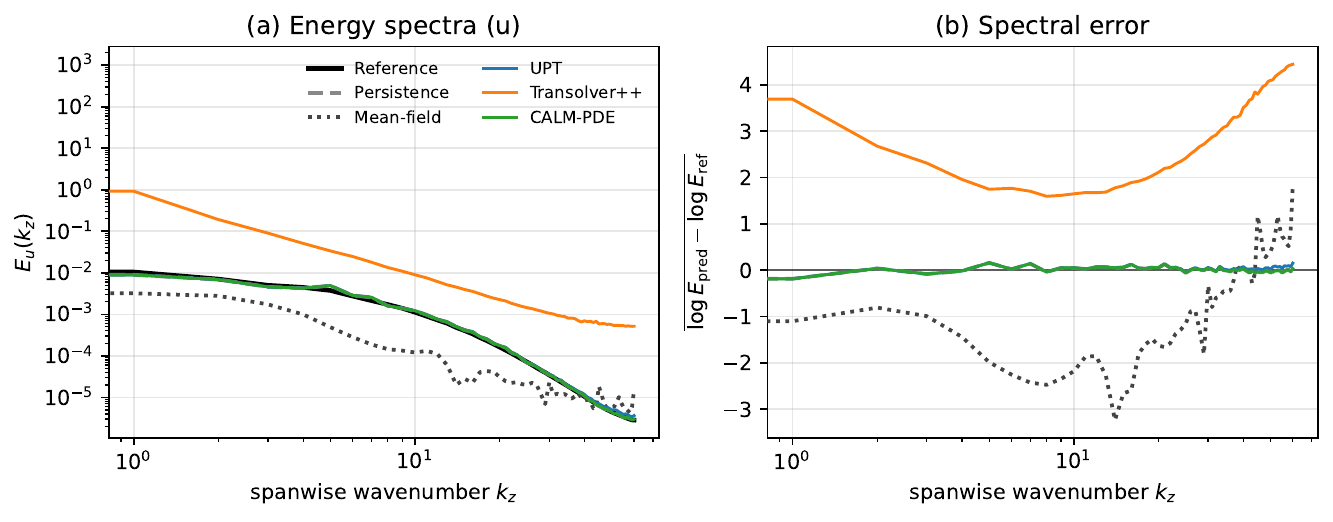}
    \caption{Spectral fidelity of autoregressive rollouts. Panel (a) shows reference and predicted \textcolor{revcolor}{spanwise} energy spectra averaged over test trajectories. Panel (b) shows the log-power discrepancy across wavenumbers.}
    \label{fig:spectral_results}
\end{figure}

\paragraph{Spectral fidelity.}
Pointwise errors do not fully reveal whether a surrogate preserves the multi-scale structure of the turbulent wake. We therefore compare \textcolor{revcolor}{spanwise} energy spectra between predicted and reference trajectories. In \autoref{fig:spectral_results}, spectra are averaged over test trajectories and rollout times, and the spectral error is shown as a function of wavenumber. A model that tracks large coherent structures but dissipates small eddies is expected to match low wavenumbers while underestimating high-wavenumber energy. We also report a persistence baseline, defined by the one-step rule \(\hat{u}_{t+1}=u_t\); under autoregressive rollout, this corresponds to repeating the initial state. Because this baseline preserves a real turbulent snapshot, it can match the time-averaged energy spectrum reasonably well despite not predicting the temporal evolution of the wake. This highlights that spectral agreement alone is insufficient and should be interpreted together with field-level and long-horizon rollout errors. Among the learned models, CALM-PDE most closely matches the reference spectrum across the resolved range, with near-zero log-power discrepancy over low and intermediate wavenumbers. UPT follows the correct spectral decay but shows larger high-wavenumber errors, indicating reduced fidelity at small scales. In contrast, Transolver++ strongly overpredicts spectral energy, especially at intermediate and high wavenumbers, suggesting spurious small-scale content during rollout.

\paragraph{Latency.}
Finally, we compare the full-mesh inference cost of the neural surrogates in the last column of \autoref{tab:temporal_stability}\textcolor{revcolor}{, together with the cost of generating one trajectory with the solver}. All evaluated methods remain costly at this spatial resolution, requiring hundreds to thousands of seconds to predict a full trajectory. UPT is the fastest baseline, followed by CALM-PDE and Transolver++, while GINO is substantially slower than the other methods\textcolor{revcolor}{, which also constrains its accuracy: it is the only baseline that did not converge within the shared training budget.} \textcolor{revcolor}{In GPU-seconds, every surrogate provides a net speedup over the solver, ranging from ${\sim}3\times$ for GINO to ${\sim}16\times$ for UPT. This margin is modest compared with the speedups commonly reported for neural surrogates on smaller problems.} These results show that, even when surrogate models improve rollout accuracy, scaling inference to dense irregular meshes with millions of points remains a major computational bottleneck.

Together, these evaluations expose complementary failure modes of neural PDE surrogates on \textsc{TurtLES}: short-horizon prediction error, long-horizon instability, loss of high-frequency turbulent energy, and insufficient computational acceleration on dense irregular meshes.

{\color{revcolor}
\paragraph{Limitations.}
\textsc{TurtLES} is restricted to one physical regime: turbulent cylinder wakes in the shear-layer transition range. Whether the failure modes observed here transfer to other geometries or Reynolds regimes remains an open question. The reference trajectories are themselves an approximation: computed via implicit-LES, so every surrogate error reported here is measured against this solver rather than against a DNS or experimental ground truth. To keep the dataset tractable, snapshots are subsampled in time; although the sampling interval is chosen to preserve the inertial range, some high-frequency information is inevitably discarded. The resulting short sampling window also rules out converged averaged statistics such as Reynolds stresses, which would require substantially longer trajectories than storage permits, limiting the physical diagnostics available for evaluation. Finally, the four baselines surveyed here are not an exhaustive account of how neural operators perform on this class of flows; building that account is the role we intend \textsc{TurtLES} to serve.}


\section{Conclusion}

We introduced \textsc{TurtLES}, a large-scale benchmark for learning neural PDE surrogates in realistic three-dimensional turbulent flow regimes. The dataset contains \textcolor{revcolor}{400} high-fidelity iLES trajectories of cylinder wakes across variations in geometry, Reynolds number, and angle of attack, with millions of irregularly distributed solution points per frame and hundreds of time steps per trajectory. By combining dense unstructured spatial discretizations, long temporal horizons, and physically turbulent wake dynamics, \textsc{TurtLES} targets a regime that is not jointly covered by existing CFD benchmarks.

Our benchmark experiments show that current neural operators face major challenges in this setting. While several models achieve reasonable short-horizon accuracy, long autoregressive rollouts reveal substantial differences in stability. Spectral diagnostics show that pointwise errors alone are insufficient: models may appear accurate while misrepresenting turbulent energy at fine scales. 

\textsc{TurtLES} is intended to support progress toward neural surrogates that are not only accurate on average, but also stable over long horizons, faithful to multiscale turbulent structure, and efficient on \textcolor{revcolor}{meshes of industrially relevant size}. Future work may extend the benchmark to additional geometries, higher Reynolds-number regimes, richer boundary conditions, and uncertainty-aware or generative surrogate models. We hope that \textsc{TurtLES} will provide a useful testbed for developing and evaluating next-generation neural operators for high-resolution computational fluid dynamics.

\begin{ack}
We acknowledge the financial support provided by DEEPNUM (ANR-21-CE23-0017-02), PHLUSIM (ANR-23-CE23-0025-02), MELISSA (ANR-24-CE23-7140-01), PEPR Sharp (ANR-23-PEIA0008, ANR, FRANCE 2030), and  cluster IA  PostGenAI@Paris (ANR-23-IACL-0007).

This project was provided with computer and storage resources
by GENCI at IDRIS thanks to the grants 2026-AD011016890R2, 2026-AD011017368, 2025-A0191016890, 2026-A0211016890 on the supercomputer Jean Zay’s V100/A100/H100 partitions.

The authors declare no competing interests.
\end{ack}

\clearpage
\bibliography{bibliography}
\bibliographystyle{plainnat}

\clearpage
\appendix


\section{Broader impact}                                
\label{app:impact}                                     
\textsc{TurtLES} provides high-fidelity three-dimensional turbulent flow trajectories on dense irregular meshes, and is intended to serve as a benchmark for the development of neural surrogate models in computational fluid dynamics. We anticipate the following uses:         
\begin{enumerate} 
      \item evaluating geometry-aware operator-learning models on dynamic, physically turbulent three-dimensional flow rather than on regular-grid or two-dimensional benchmarks;           
      \item studying the long-horizon stability and multiscale spectral fidelity of learned PDE surrogates through the evaluation protocol introduced in \autoref{sec:results};   
      \item informing the design of new architectures for irregular meshes with millions of degrees of freedom, where memory and inference cost become first-order constraints;
      \item extending point-cloud and graph-based benchmarks toward industrially relevant fluid-dynamics regimes;
      \item supporting the development of surrogate solvers that accelerate early-stage design exploration in aerospace and automotive engineering;
      \item studying the generalization behavior of neural operators across geometry, Reynolds number, and angle of attack in a controlled parametric setting.
  \end{enumerate}

We do not anticipate direct negative societal impact: the dataset is synthetic and contains no personally identifiable, geographic, or sensitive information. As with any data-driven surrogate, downstream practitioners should validate predictions before relying on them for safety-critical engineering decisions; surrogate models trained on \textsc{TurtLES} are not substitutes for verified high-fidelity simulation in such contexts.


\section{Reproducibility statement} \label{app:repro}
\textcolor{revcolor}{The project page provides} (i)~the code used to generate every table and figure in the main paper; and (ii)~\textcolor{revcolor}{a link to the dataset hosted on Hugging Face,} along with instructions for re-running the data generation pipeline (mesh generation with Gmsh, simulation with PyFR, post-processing into HDF5).

All baseline experiments were conducted on a single NVIDIA H100~80~GB GPU per run, with approximately 20~hours of wall-clock training time per model. Evaluation runs use the same hardware. Hyperparameters, data subsampling, optimization schedule, and inference protocol are documented in \autoref{app:baselines}.

 
\section{Description of software} \label{app:software}   
This work uses computational fluid dynamics, mesh-generation, and machine learning tools, all of them open source.
      
\paragraph{PyFR~\citep{WITHERDEN20143028}.} PyFR is an open-source GPU-accelerated solver for advection--diffusion-type problems, based on the Flux Reconstruction high-order accurate scheme. We use PyFR for all iLES simulations. We used version~\textcolor{revcolor}{2.1} of PyFR with CUDA~12 on NVIDIA H100 GPUs. PyFR is released under the BSD 3-Clause license. 

\paragraph{Gmsh~\citep{GMSH}.} Gmsh is an open-source three-dimensional finite-element mesh generator with a built-in CAD engine and post-processor. We use Gmsh, scripted from Python, to build all \textcolor{revcolor}{400} meshes in the dataset (block-structured near-wall regions, unstructured wake regions, structured spanwise extrusion). We used version~4.11. Gmsh is released under the GNU GPL.   

\paragraph{HDF5 / h5py~\citep{hdf5}.} The dataset is stored in HDF5 chunks (one coordinate file plus four temporal chunks of 100 frames per trajectory) and read at training and evaluation time through h5py. HDF5 is released under a permissive BSD-style license; h5py is released under the BSD 3-Clause license.                              
\paragraph{PyVista~\citep{pyvista}.} PyVista is an open-source tool for 3D visualization and mesh analysis built on top of the Visualization Toolkit (VTK). In this work, we use it to post-process the physical fields produced by our simulations. PyVista is released as free and open-source software under the MIT License.
  
\paragraph{ParaView~\citep{ParaView}.} We use ParaView for qualitative visualization of the simulated and predicted flow fields. We used version~5.13. ParaView is released under the BSD 3-Clause license. 

\paragraph{PyTorch~\citep{Ansel_PyTorch_2_Faster_2024}.} All neural-network training and inference is implemented in PyTorch. We used PyTorch~2.8 with CUDA~12.8 and Python~3.12. PyTorch is released under the BSD 3-Clause license.                                     

\paragraph{neuraloperator~\citep{kossaifi2025librarylearningneuraloperators}.} The GINO baseline is built on the official \texttt{neuraloperator} library, which provides the GNO and FNO building blocks used in the encoder--FNO--decoder pipeline described in \autoref{app:baselines}. We used version~1.0.2. \texttt{neuraloperator} is released under the MIT license.


\section{Constant and Dimensionless Quantities} \label{sec:appquan}
In compressible flow simulations, the dynamics are primarily governed by the Reynolds number ($Re$), Mach number ($Ma$) and Prandtl number ($Pr$). In addition, cylinder flows exhibit periodic unsteady behavior which warrants the introduction of a non-dimensional frequency, namely the Strouhal number ($St$). We mathematically define these quantities as follows:
\begin{equation}
    Re = \frac{\rho U D}{\mu}, \quad
    Ma = \frac{U}{c}, \quad
    Pr = \frac{\mu c_p}{k}, \quad
    St = \frac{f D}{U}.
\end{equation}
$U$ and $D$ denote the characteristic velocity and length scale of the problem, respectively. In the present study, $U$ is equal to the inflow velocity $U_\infty$, and \textcolor{revcolor}{$D = 2b = 1$ is the object thickness defined in \autoref{sec:appsample}}. Furthermore, $\rho$ is the fluid density, $
\mu$ the dynamic viscosity, $c$ the speed of sound, $c_p$ the specific heat, $k$ the thermal conductivity, and $f$ is a characteristic frequency associated with the flow. For all flow cases considered in this work, the Mach number $Ma = 0.2$ and Prandtl number $Pr = 0.71$ are kept constant while the Reynolds number is varied. Finally, time $t$ is always presented in dimensionless form, normalized by the characteristic time scale $D/U$.


\section{Implicit Large Eddy Simulation} \label{sec:appiLES}
The Navier--Stokes equations represent the fundamental laws governing fluid dynamics. For a Cartesian coordinate system, \autoref{eq:NSmass}, \autoref{eq:NSmomentum}, and \autoref{eq:NSenergy} respectively denote a fluid's conservation of mass, momentum and energy:
\begin{align}
    \frac{\partial \rho}{\partial t} + \frac{\partial}{\partial x_i}(\rho u_i) &= 0, \label{eq:NSmass} \\[10pt]
    \frac{\partial}{\partial t} \rho u_i + \frac{\partial}{\partial x_j}(\rho u_i u_j) &= -\frac{\partial p}{\partial x_i} + \frac{\partial}{\partial x_j} \tau_{ij}, \label{eq:NSmomentum} \\[10pt]
    \frac{\partial}{\partial t} E + \frac{\partial}{\partial x_j}(E u_j) &= \frac{\partial}{\partial x_j} \left[ u_i \tau_{ij} - q_j - p u_j \right], \label{eq:NSenergy}
\end{align}
where $\rho$ is the fluid density, $E$ is the total energy per unit volume and $u_i$ denotes the velocity in the $x_i$ direction (with $i = 1,2,3$ corresponding to the $x, y, z$ axes). Note that these equations utilize Einstein notation where a repeated index ($i$, $j$, $k$) implies a summation over all spatial dimensions. Furthermore, the stress tensor, strain tensor and heat flux are defined as: 
\begin{equation}
    \tau_{ij} = 2\mu \left[ S_{ij} - \frac{1}{3} \frac{\partial u_k}{\partial x_k} \delta_{ij} \right], \qquad
    S_{ij} = \frac{1}{2} \left( \frac{\partial u_i}{\partial x_j} + \frac{\partial u_j}{\partial x_i} \right), \qquad
    q_j = -\frac{\mu}{Pr} \frac{\partial h}{\partial x_j}.
    \label{eq:constitutive_relations}
\end{equation}
Here, $\mu$ is the fluid's dynamic viscosity, $Pr$ is the Prandtl number, and $h$ is the specific enthalpy. Finally, the ideal gas equation of state provides a relation coupling the pressure to the total energy:
\begin{equation}
    p = (\gamma - 1) (E - \frac{1}{2} \rho u_i u_i),
\end{equation}
where $\gamma$ is the specific heat ratio. An analytical solution to this system of equations only exists under simplifying assumptions, and thus numerical methods are used to approximate solutions for complex flow problems. The domain is then discretized into a grid on which flow variables are iteratively solved. To capture the entire turbulence spectrum of a fluid, Direct Numerical Simulation (DNS) employs a grid fine enough to resolve the smallest physical structures. The computational overhead is often intractable for industrially relevant applications, as the cost of fully resolving all turbulent scales in 3D Navier-Stokes simulations grows with $Re^3$, where $Re$ denotes the Reynolds number. Instead, Reynolds-averaged Navier-Stokes (RANS) based simulations decompose an instantaneous fluid quantity into its time-averaged and fluctuating component, whereby the effects of the latter are represented by a turbulence model rather than computed directly. While this allows a significant reduction in computational cost, these methods exhibit limited accuracy for complex flow phenomena (e.g. laminar-to-turbulent transition, separation, three-dimensional effects) \citep{wilcox1998turbulence}.

Large Eddy Simulation (LES) provides a high-fidelity alternative for simulating turbulent flows at reasonable cost. It is based on the separation between the energy-containing large scales of turbulence and the smaller high-frequency scales through a filter operation \citep{PopeBook}:
\begin{equation} \label{eq:fildeco}
    u_i(x, t) = \hat{u}_i(x,t) + u_i''(x,t).
\end{equation}  
The filtered flow variables ($\hat{u}_i$) are also referred to as the resolved variables, as they are directly solved for in the LES framework. The residual of the filter operation, $u_i''$, represents the unresolved turbulent motions. Numerous filtering strategies exist, but often the computational mesh itself acts as a filter where the local grid size, $\Delta_{grid}$, dictates the cut-off between represented turbulent scales and \textit{sub-grid scales} (SGS). The scale separation of \autoref{eq:fildeco} is physically motivated by each component's distinct behavior. The resolved scales are highly energetic, anisotropic, and strongly influenced by the flow geometry, which requires their direct computation. In contrast, the unresolved scales are more isotropic, carry less energy and exhibit a more universal behavior, which allows for unified approaches to modeling their effects. Because these smaller scales are not resolved, LES utilizes significantly coarser meshes than DNS, leading to reduced computational complexity.

In explicit LES, an explicit model is used to replicate the effects of SGS turbulence, which entails providing the correct energy drain from the resolved flow to the smaller unresolved scales. This approach assumes that numerical errors associated with the discretization scheme are small. In practice, this is often not the case, as even high-order schemes exhibit truncation errors of similar magnitude to SGS model contributions \citep{GHOSAL1996187}. Both truncation errors and SGS model terms are typically most significant at the smallest resolved scales. Due to the strong non-linear nature of Navier-Stokes, their combined effect can be difficult to anticipate. One way to address this challenge is to apply an explicit filter at a scale larger than the grid spacing ($\Delta_{filter} > \Delta_{grid}$), effectively shifting the SGS model's activity away from the high-frequency truncation errors. However, this significantly increases computational cost as much finer grids are required to maintain the same level of resolved turbulence. 

Alternatively, by recognizing that some numerical schemes exhibit errors of similar dissipative nature as the effect of the sub-grid scales, implicit LES (iLES) abandons the use of an explicit SGS model and lets the truncation error serve as the energy drain from the resolved flow. The present study adopts this approach by employing the Flux Reconstruction (FR) or Correction Procedure via Reconstruction (CPR) scheme \citep{Huynh2007}. CPR functions well in the context of iLES as the scheme is inherently dissipative at high wave numbers \citep{VINCENT20118134}.


\section{Simulation setup} \label{sec:appsimset}

\begin{figure}[t]
    \centering
    \begin{minipage}[b]{0.45\textwidth}
        \centering
        \includegraphics[width=0.95\textwidth]{figures/setup/Domain3D_edit.png}
        \\ (a) 
    \end{minipage}
    \hfill
    \begin{minipage}[b]{0.52\textwidth}
        \centering
        \includegraphics[width=0.95\textwidth]{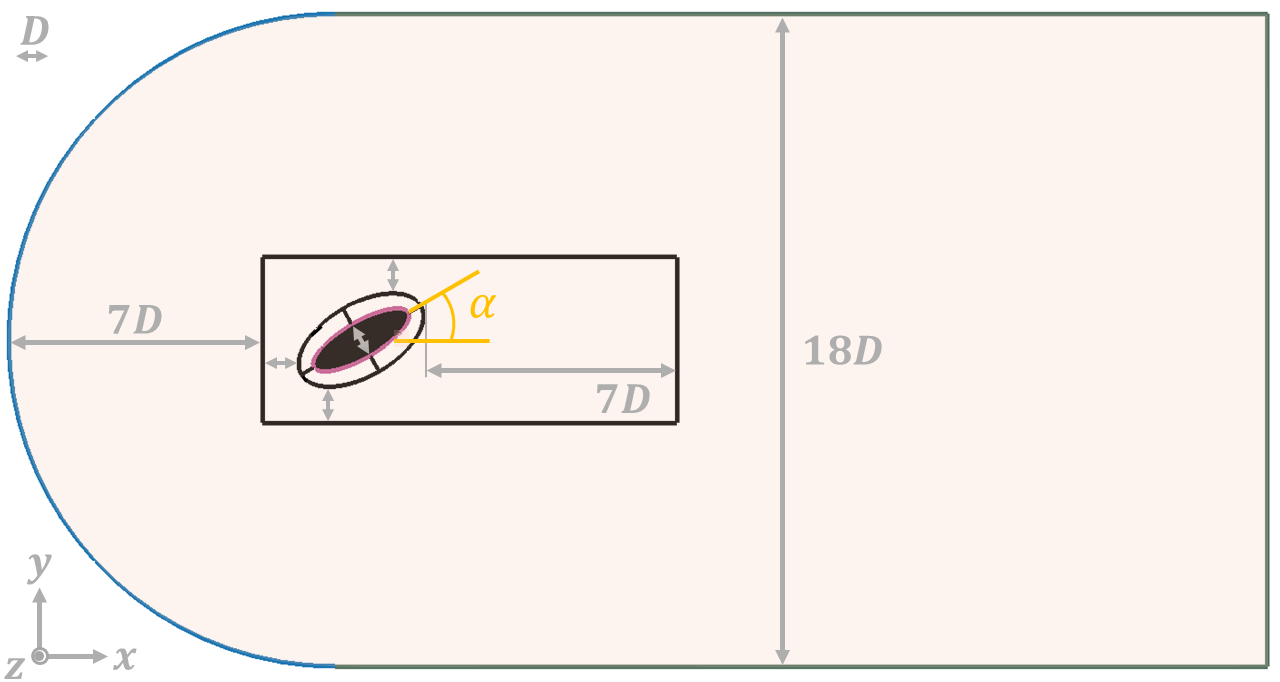}
        \\ (b) 
    \end{minipage}
    \caption{Schematic overview of the computational domain. (a) Three-dimensional view illustrating the physical boundaries, including the inlet (\textbf{A}), outlets (\textbf{B}--\textbf{D}), object surface (\textbf{E}), and periodic side-walls (\textbf{F}, \textbf{G}). (b) 2D side view showing characteristic dimensions relative to the \textcolor{revcolor}{object thickness} $D$. Note that black lines have no physical implications but indicate regions used for mesh refinement.}
    \label{fig:Domains}
\end{figure}

\paragraph{Domain and Boundary Conditions}
The compressible Navier-Stokes equations are solved within the three-dimensional domain in \subfigref{fig:Domains}{a}. This domain is generated by extruding the 2D plane depicted in \subfigref{fig:Domains}{b} by a spanwise length of $2\pi$. The center of the super-elliptical object $\textbf{E}$ (colored in pink) is placed at the origin $(x,y) = (0,0)$ and rotated according to its prescribed inflow angle $\alpha$. Rotating the object, rather than modifying the inflow angle of the flow field, is more convenient as it ensures the wake always propagates in the horizontal direction, which simplifies meshing and deciding what solution region to extract during data processing. The five blocks defined by the black lines in \subfigref{fig:Domains}{b} have no physical meaning but are used to control the automated meshing. The four blocks directly surrounding the physical object define the structured meshing region by placing a super-ellipse with $a'$, $b'$ and $n'$ (see \autoref{sec:appsample}) and connecting it to  object $\textbf{E}$ with parameters $a$, $b$ and $n$. In all cases, $a'= a + D/2$ and $b'= b + D/2$ in order to create a thick enough structured mesh region that can capture the initial development of separated shear layers. $n'$ is discretely defined as $1.5, 2, 5,$ and $10$ for $n = 1, 2, 5,$ and~$\infty$, respectively. These values mitigate mesh skewness, which can otherwise lead to numerical issues. The rectangular box is used to control the mesh nearby and behind the object, where a fine resolution is required to capture complex wake structures. Its front, top and bottom boundaries are located $1D$ from the structured mesh region, while the back edge extends to $7D$ behind the super-ellipses. Finally, the inlet $\textbf{A}$ is positioned at $7D$ ahead of the rectangular wake box while the outlet curves \textbf{B}--\textbf{D} always extend until $x/D = 25$. \autoref{tab:BCs} provides an overview of the boundary conditions placed on the domain.

\begin{table}[t]
  \caption{Boundary conditions of the simulation. Surface symbols refer to boundaries indicated in \subfigref{fig:Domains}{a}}
  \label{tab:BCs}
  \centering
  \begin{tabular}{lll}
    \toprule 
    Category  & Symbol (in \subfigref{fig:Domains}{a})  & Boundary condition \\
    \midrule
    Inlet & $\textbf{A}$ & Riemann invariant    \\
    Outlet     & \textbf{B}--\textbf{D} & Riemann invariant      \\
    Wall     &  $\textbf{E}$     & Adiabatic no-slip wall  \\
    Left side   &   $\textbf{F}$     & Periodic  \\
    Right side   &   $\textbf{G}$     & Periodic  \\

    \bottomrule
  \end{tabular}
\end{table}

\paragraph{Meshing}
Generating a high-quality mesh is essential for achieving accurate and stable CFD simulations. Key considerations include minimizing element distortions, maintaining low skewness, ensuring gradual transitions in cell size, and providing adequate resolution. The latter is of particular importance in implicit large eddy simulations, where it is assumed that a significant portion of turbulence scales is directly resolved. In practice, a high-quality mesh is attained through case-specific iterative refinement, which is not feasible when constructing a large dataset. It was therefore chosen to adopt a simple meshing strategy that can be automated, accommodates a wide range of flow cases and limits the number of variable parameters. The routine is implemented using Python and Gmsh \citep{GMSH}, an open source three-dimensional finite element mesh generator. Gmsh supports the generation of high-order meshes, which improve accuracy for domains with curved boundaries.

\begin{figure}[t]
    \centering
    \begin{minipage}[b]{0.55\textwidth}
        \centering
        \includegraphics[width=0.95\textwidth]{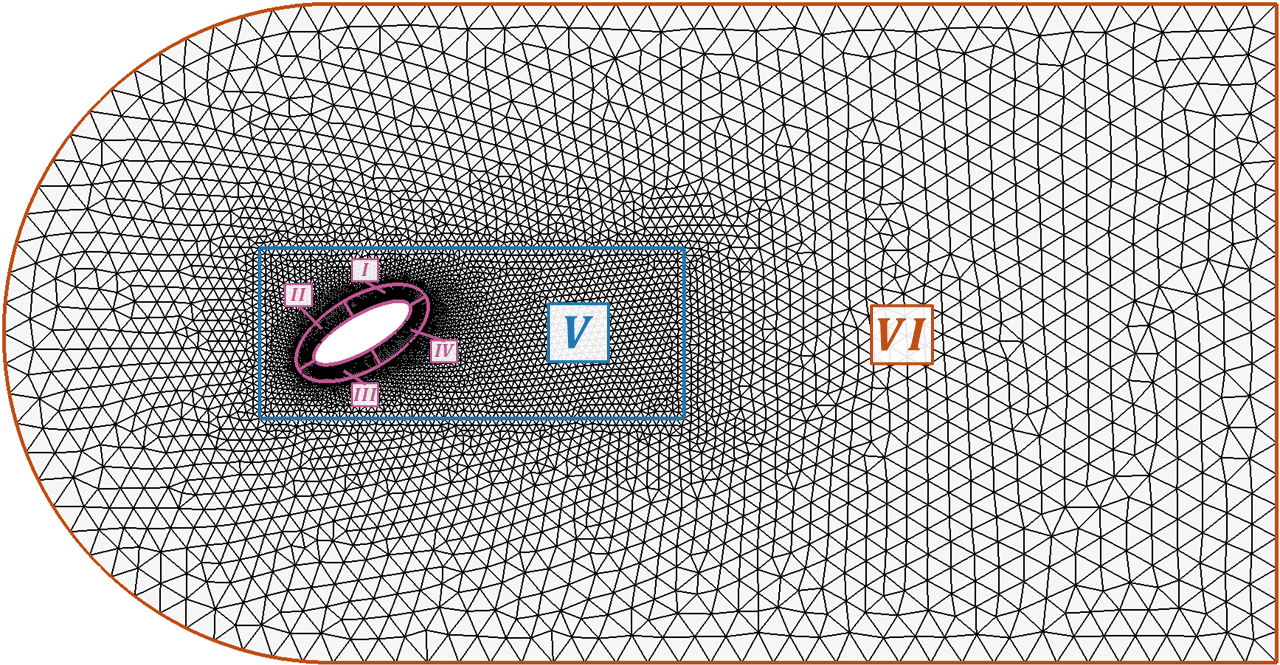}
        \\ (a) 
    \end{minipage}
    \hfill
    \begin{minipage}[b]{0.42\textwidth}
        \centering
        \includegraphics[width=0.95\textwidth]{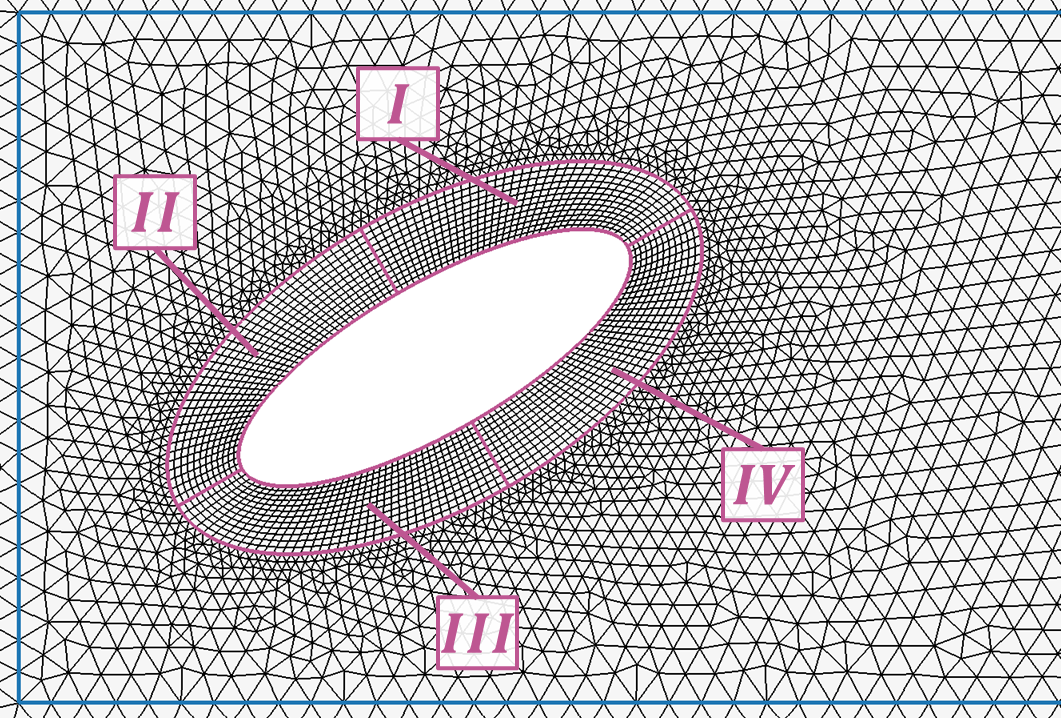}
        \\ (b) 
    \end{minipage}
    \caption{Two-dimensional ($xy$-)plane of the computational mesh. The full three-dimensional domain is a 30-layer extrusion in the $z$ direction. (a) Full domain. (b) Zoomed-in view of the geometric object.}
    \label{fig:meshes}
\end{figure}

The $xy$-plane is partitioned into six blocks as illustrated in \autoref{fig:meshes}. The dimensions of each block are defined in \subfigref{fig:Domains}{b} and remain consistent across all flow cases. Blocks $\textbf{V}$ and $\textbf{VI}$ are discretized using unstructured triangular elements, as they provide flexibility in dealing with geometric variations. For each block, a target cell size is set: $l_{wb}$ (for $\textbf{V}$) and $l_{ff}$ (for $\textbf{VI}$). Gmsh targets these sizes when possible while ensuring smooth transitions between regions of differing resolution. Blocks $\textbf{I}$-$\textbf{IV}$ employ structured layers to give precise control over the near-wall discretization. Within these regions, the first element thickness $l_t$ is chosen such that $y^+<1$. All other wall-normal cell sizes are determined through a geometric progression with a growth rate of 1.021. Furthermore, the cell size in the wall-parallel direction is set by $l_p$. 

Mesh resolution is highly dependent on Reynolds number; for increasing $Re$, a finer mesh is required to capture the smaller turbulent structures. To determine the appropriate values of the parameters $l_{wb}$, $l_{ff}$, $l_t$, and $l_p$, mesh convergence studies are conducted at the edge cases of $Re = 1000$ and $Re = 3900$ in \autoref{sec:appvali}. Between these Reynolds numbers, cell sizes are scaled using a power law interpolation. This resulting two-dimensional mesh is extended into three dimensions by a 30-layer structured extrusion in the span-wise direction. 

\section{Validation} \label{sec:appvali}
\begin{figure}[t]
    \centering
    \includegraphics[width=\linewidth]{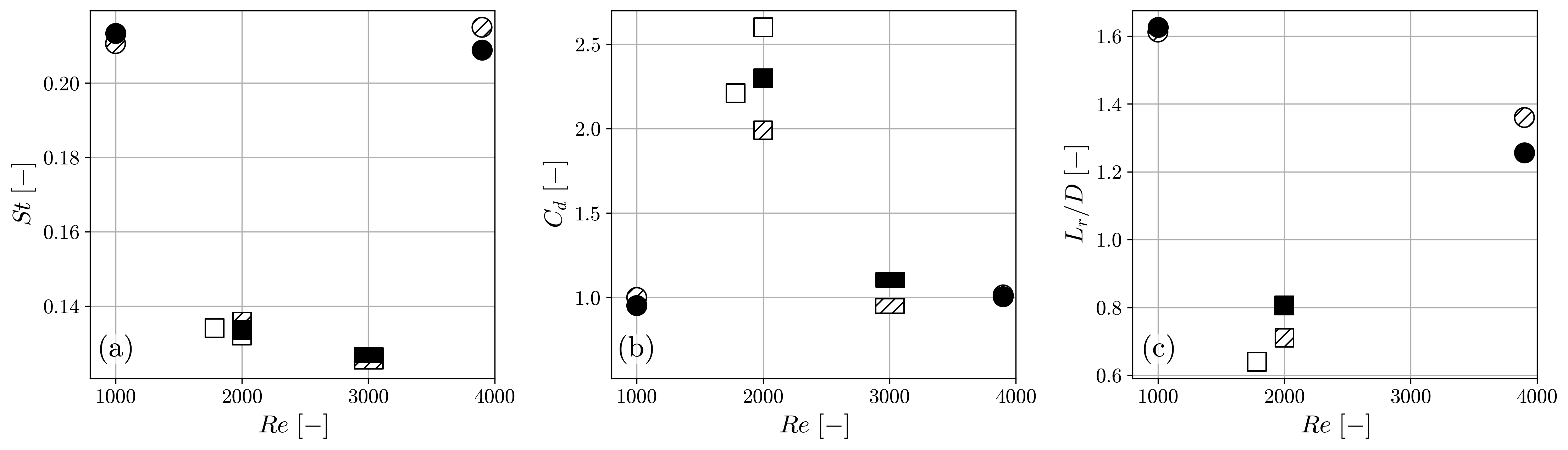}
    \caption{Comparison between present simulations (solid black markers) and validation data from the literature (open and hatched markers) for (a) Strouhal number, (b) drag coefficient, and (c) recirculation length. Validation data are taken at discrete Reynolds numbers from the following sources: Re=1000 \citep{Jiang_Cheng_2017}; Re=1783 \citep{Bai2018}; Re=2000 (hatched) \citep{sharma_flow_2022} and (open) \citep{Brun2008}; Re=3000 \citep{cimarelli_direct_2018}; Re=3900 \citep{lehmkuhl_low-frequency_2013}.}
    \label{fig:validationstats}
\end{figure}

The simulation is validated using DNS and experimental data of four benchmark cases: a circular cylinder flow ($a = 0.5$, $n = 2$) at $Re  = 1000$ and $Re = 3900$, an elongated rectangle ($a = 2.5$, $n = \infty$) at $Re = 3000$ and a square rectangle ($a = 0.5$, $n = \infty$) at $Re = 2000$. These cases were selected to span a broad range of Reynolds numbers and to assess performance on sharp-edged and rounded geometries with varying aspect ratios. \autoref{fig:validationstats} presents the overall statistics for each case together with reference data from literature. In general, the simulation shows good agreement with DNS and experimental results. Discrepancies are observed in drag prediction for very bluff bodies, but this appears to be a common trend for the studies included in the figure. These cases are discussed in detail below, with the cylinder flow cases also being used for a mesh convergence study.

\subsection*{Circular cylinder ($a = 0.5$, $n = 2$) at $Re  = 1000$ and $Re = 3900$}
The circular cylinder flow case is used for a mesh convergence study, ensuring the numerical results are independent of spatial discretization. Simulations are conducted for 1500 non-dimensional time units on increasingly refined grids and compared against DNS data from \citet{lehmkuhl_low-frequency_2013}. Refinements are achieved by adjusting the parameters $l_{wb}$, $l_{ff}$, $l_t$ and $l_p$ described in \autoref{sec:appsimset}, yielding computational meshes with 12.9M, 20.0M, 32.6M solution points \textcolor{revcolor}{in the full computational domain}, hereafter referred to as the coarse, medium and fine grids, respectively. \textcolor{revcolor}{Note that the dataset only stores the solution points inside the extraction box defined in the data processing paragraph of \autoref{sec:appsample}, so the per-case point counts reported there are lower.}

The flow around a circular cylinder at $Re = 3900$ has been studied extensively, yet varying results are reported in the literature. \citet{lehmkuhl_low-frequency_2013} addressed these inconsistencies by demonstrating that the flow exhibits bistable behavior, oscillating between two modes, characterized by the length of the separated wake region, also referred to as the \textit{recirculation length}. Previous studies relied on observation windows that were too short to record both modes, leading to results that reflected either the short-bubble state \citep{Loureno1993CharacteristicsOT} or the long-bubble state \citep{parnaudeau_experimental_2008}.

\begin{figure}[t]
    \centering
    \includegraphics[width=0.95\linewidth]{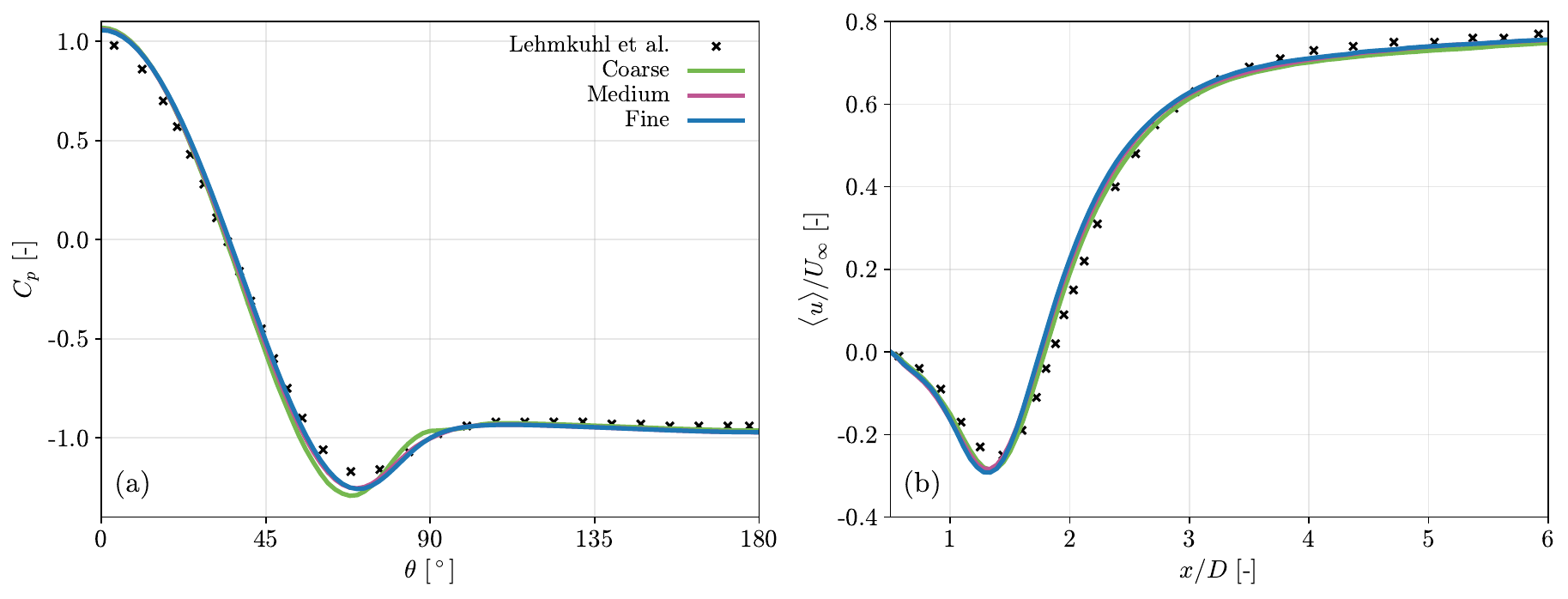}
    \caption{Results for the coarse, medium and fine computational grids: (a) pressure coefficient distribution over the cylinder surface as a function of the angular position $\theta$, defined from the front stagnation point ($\theta=0^\circ$) and increasing along the cylinder surface in the clockwise direction; and (b) mean streamwise velocity along the centerline ($y/D=0$) downstream of the cylinder. Validation data are taken from \citet{lehmkuhl_low-frequency_2013}.}
    \label{fig:cylpeanu}
\end{figure}

\subfigref{fig:cylpeanu}{b} displays the streamwise velocity along the centerline ($y/D=0$) downstream of the cylinder. The three mesh resolutions yield nearly identical profiles in agreement with the DNS data, although a slightly shorter recirculation bubble is consistently observed. \subfigref{fig:cylpeanu}{a} shows the pressure coefficient distribution over the cylinder surface, defined as $C_p = (p - p_\infty) / \frac{1}{2}\rho_\infty U_\infty^2$, where $\infty$ denotes free-stream conditions. Due to symmetry, the results from the upper and lower surfaces are averaged together. At $\theta=0^\circ$, the flow encounters the stagnation point where its velocity reduces to zero and $C_p$ attains its maximum. For incompressible flow, as in \citet{lehmkuhl_low-frequency_2013}, $C_p = 1$ is expected when the flow velocity is zero, consistent with Bernoulli's principle. In the present study, however, $C_p$ slightly exceeds 1. This deviation can be attributed to compressibility effects, as the flow is computed at $Ma = 0.2$. Furthermore, minor discrepancies with the reference data are observed between 45 and 90 degrees, which is consistent with solutions that capture more of the short-bubble state.

\begin{figure}[t]
    \centering
    \includegraphics[width=0.95\linewidth]{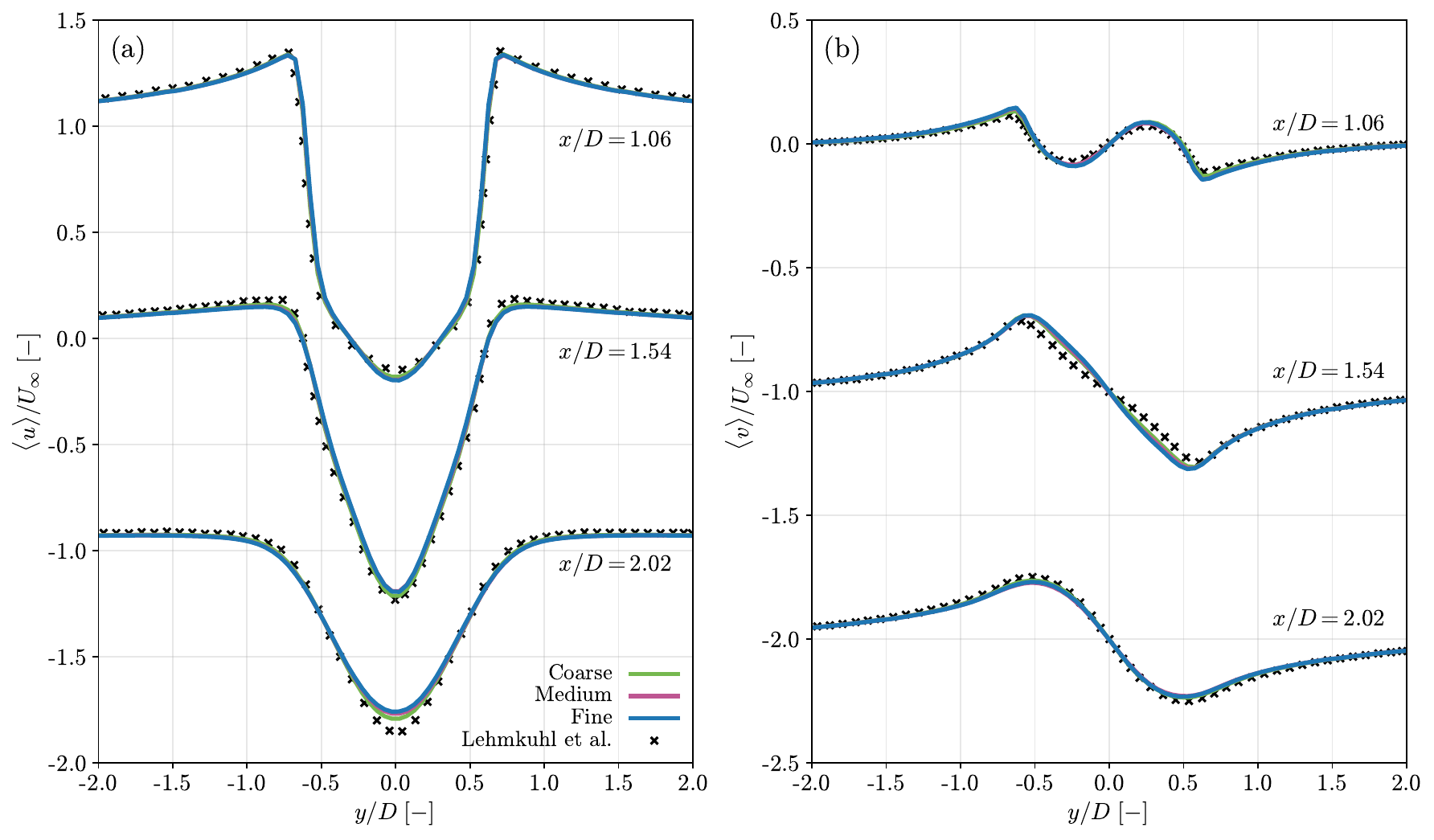}
    \caption{Results for the coarse, medium, and fine computational grids: mean wake velocity profiles of (a) streamwise component $u$ and (b) transverse component $v$ at downstream locations $x/D=1.06$, $1.54$, and $2.02$. For clarity, profiles at $x/D=1.54$ and $2.02$ are shifted downward. Validation data are taken from \citet{lehmkuhl_low-frequency_2013}.}
    \label{fig:cylwake}
\end{figure}

\begin{figure}[t]
    \centering
    \includegraphics[width=0.95\linewidth]{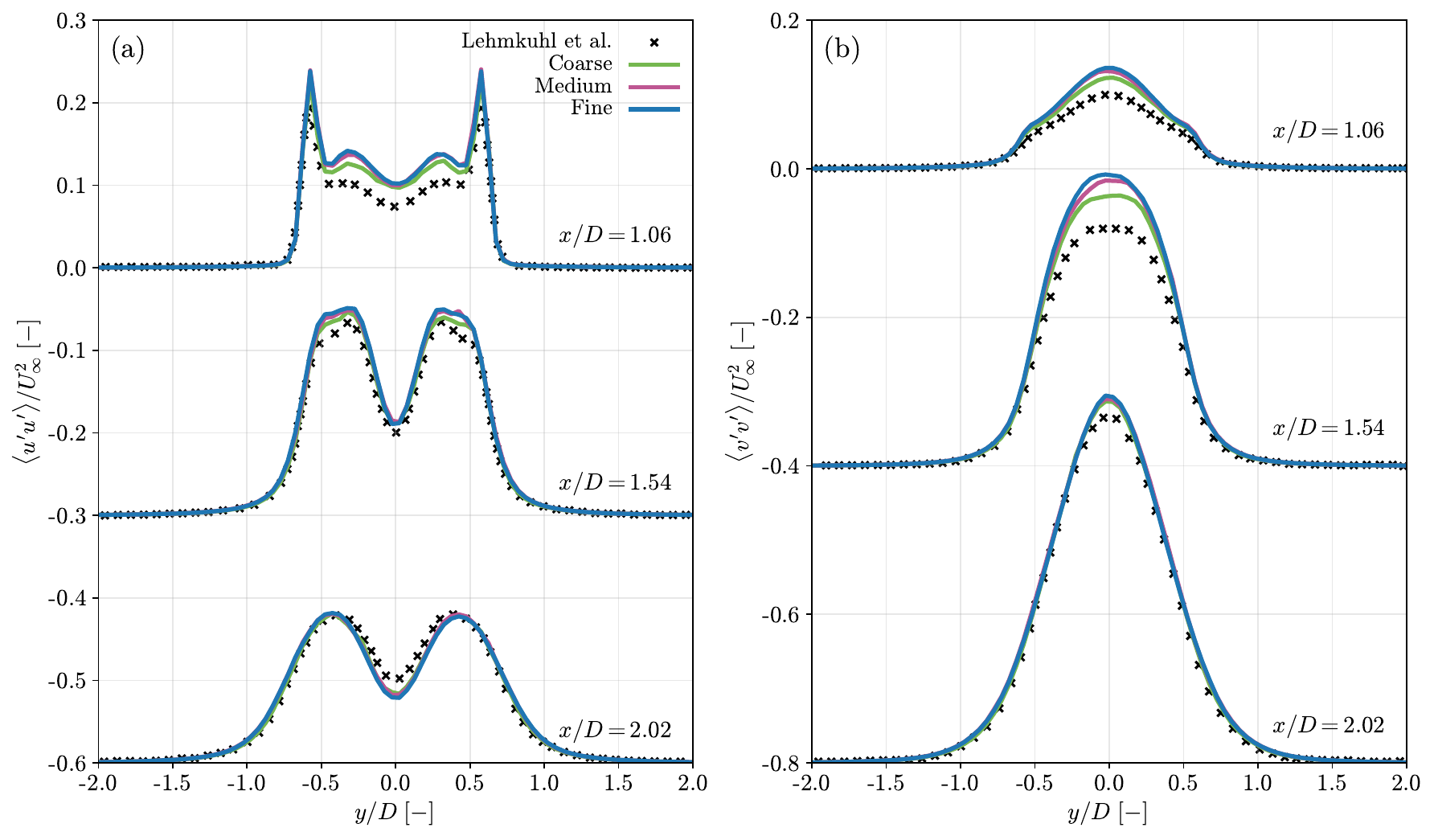} %
    \caption{Results for the coarse, medium, and fine computational grids: mean fluctuation profiles of (a) streamwise component $u'u'$ and (b) transverse component $v'v'$ at downstream locations $x/D=1.06$, $1.54$, and $2.02$. For clarity, profiles at $x/D=1.54$ and $2.02$ are shifted downward. Validation data are taken from \citet{lehmkuhl_low-frequency_2013}.}
    \label{fig:cylfluc}
\end{figure}

Time-averaged wake velocities and their fluctuations at positions $x/D = 1.06, 1.54$ and $2.02$ are presented in \autoref{fig:cylwake} and \autoref{fig:cylfluc}, respectively. Fluctuations are computed by subtracting the temporal mean from the velocity signal: $u' = u - \langle u \rangle$. The mean velocity profiles show an excellent agreement with \citet{lehmkuhl_low-frequency_2013}. The only notable discrepancy occurs for the streamwise velocity at $x/D = 2.02$ where the simulations predict a slightly faster recovery of the velocity deficit, consistent with short-bubble solutions. The fluctuation profiles exhibit larger deviations from the reference data, a challenge commonly reported in the literature. For instance, \citet{vermeire_utility_2017} observed over-predictions of similar magnitude and character when comparing second-order statistics to \citet{lehmkuhl_low-frequency_2013}.

Comparing the different computational grids shows that the medium and fine grid produce equivalent results. In contrast, the coarse grid converges to a different solution for the pressure distribution in \subfigref{fig:cylpeanu}{a} over the range $45^\circ < \theta < 90^\circ$ and across the fluctuation profiles in \autoref{fig:cylfluc}, especially for $\langle v'v'\rangle$ at $x/D=1.54$. Therefore, the medium grid is selected, as it ensures mesh-independent results at a reduced cost relative to the fine grid. 

An identical mesh convergence procedure is performed for the $Re = 1000$ case using grids with 3.7M, 7.0M and 12.9M solution points \textcolor{revcolor}{(full domain, before solution box extraction; see \textit{Data processing} in \autoref{sec:appsample})}. Due to the limited availability of reliable surface and wake validation data in the literature, detailed profile comparisons are omitted. Grid convergence is achieved on the finest mesh (12.9M points) and its statistics are compared with \citet{Jiang_Cheng_2017}. The Strouhal number, drag coefficient and recirculation length presented in \autoref{fig:validationstats} demonstrate a strong agreement. Furthermore, close agreement is found for the lift fluctuations and base pressure coefficient. The present study yields $C_{l,\mathrm{rms}} = 0.097$ and $C_{pb} = -0.8388$, compared to 0.100 and $-0.8148$, respectively, reported by \citet{Jiang_Cheng_2017}.

\begin{figure}[t]
    \centering
    \includegraphics[width=\linewidth]{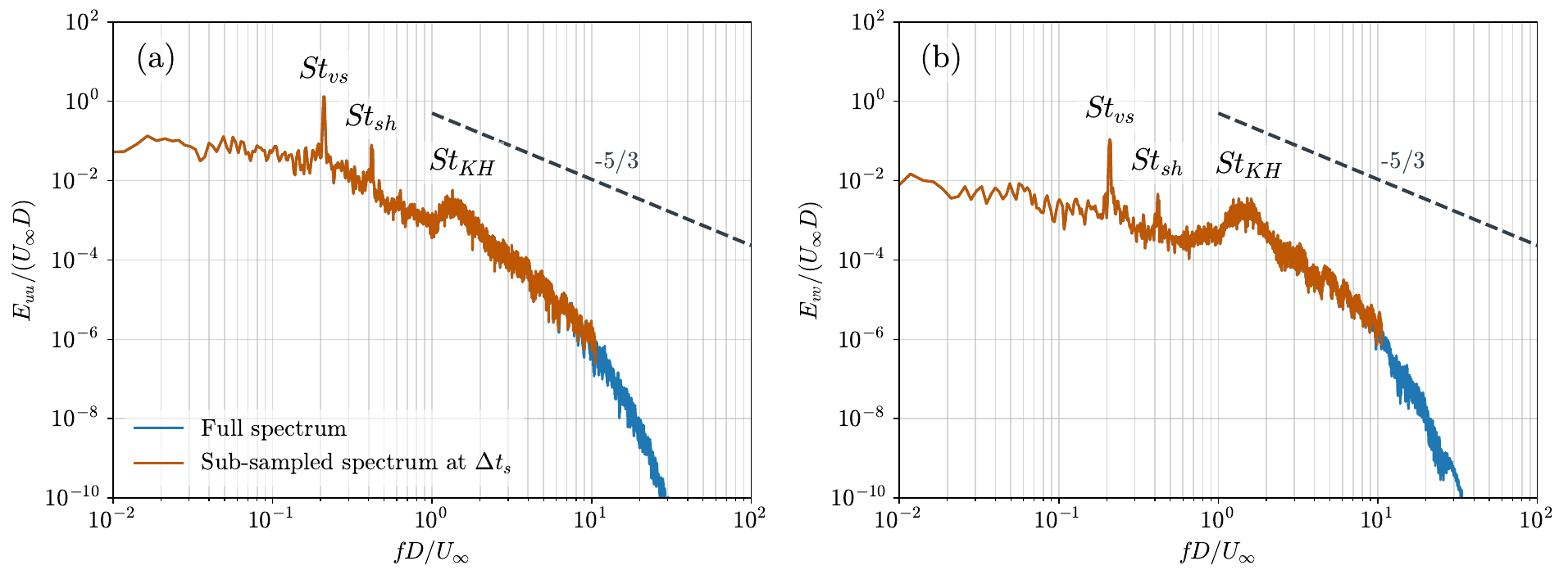}
    \caption{Comparison of velocity power spectra for (a) stream-wise and (b) cross-stream components at position $(x/D, y/D) = (0.71, 0.66)$. Spectra are shown for both the full temporal resolution and the sub-sampled resolution $\Delta t_s$. The characteristic frequencies ${St}_{vs}$, ${St}_{sh}$, and ${St}_{KH}$ are annotated for reference alongside the theoretical $-5/3$ turbulence decay slope.}
    \label{fig:cylen071}
\end{figure}

\begin{figure}[t]
    \centering
    \includegraphics[width=\linewidth]{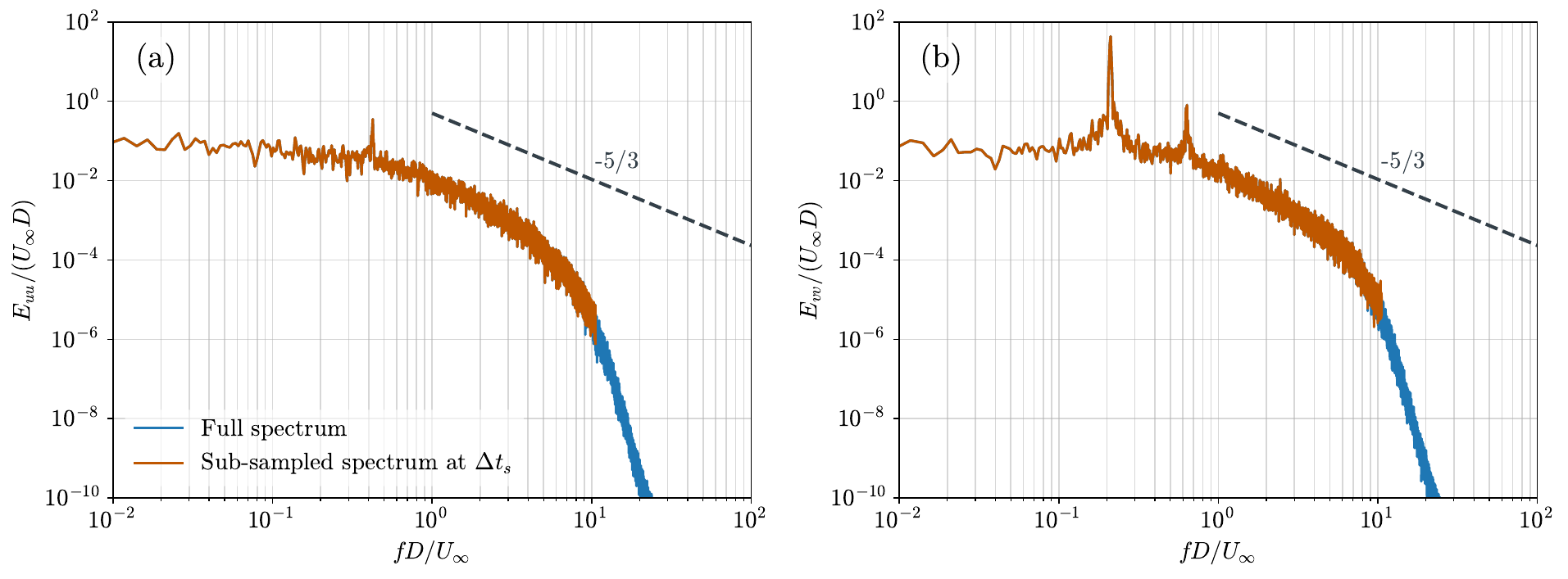}
    \caption{Comparison of velocity power spectra for (a) stream-wise and (b) cross-stream components at position $(x/D, y/D) = (3.0, 0.0)$. Spectra are shown for both the full temporal resolution and the sub-sampled resolution $\Delta t_s$. The theoretical $-5/3$ turbulence decay slope is indicated for reference.}
    \label{fig:cylen300}
\end{figure}

The Power Spectral Density (PSD) of the stream-wise and cross-stream velocity is used to further characterize the unsteady dynamics of the wake for the $Re = 3900$ case on the selected medium mesh. The PSD quantifies how the energy of velocity fluctuations is distributed across frequencies, thereby providing insight into dominant flow structures and their associated time scales. \autoref{fig:cylen071} displays the PSD in the shear layer at $(x/D, y/D) = (0.71, 0.66)$, while \autoref{fig:cylen300} shows the spectrum in the middle of the wake at $(x/D, y/D) = (3.0, 0.0)$. All subsequent discussions are based on the full spectrum result (blue line). The sub-sampled spectrum (orange line) is addressed separately in \autoref{sec:appproc}, where the information loss caused by sub-sampling the solution, necessitated by storage constraints, is examined. Overall, the spectra are in good agreement with \citet{lehmkuhl_low-frequency_2013}. The spectra exhibit the expected energy decay from low to high frequency scales, with an inertial range scaling in accordance with the theoretical $-5/3$ Kolmogorov law. Distinct peaks associated with characteristic flow phenomena are indicated in \autoref{fig:cylen071}. The vortex shedding peak at ${St}_{vs}$ represents the periodic release of vortices from the cylinder surface, while ${St}_{sh} = 2{St}_{vs}$ denotes the second harmonic of this process. Finally, the Kelvin-Helmholtz instability within the shear layer is observed at ${St}_{KH}$.

\subsection*{Elongated rectangle ($a = 2.5$, $n = \infty$) at $Re = 3000$}
\begin{figure}[t]
    \centering
    \includegraphics[width=\linewidth]{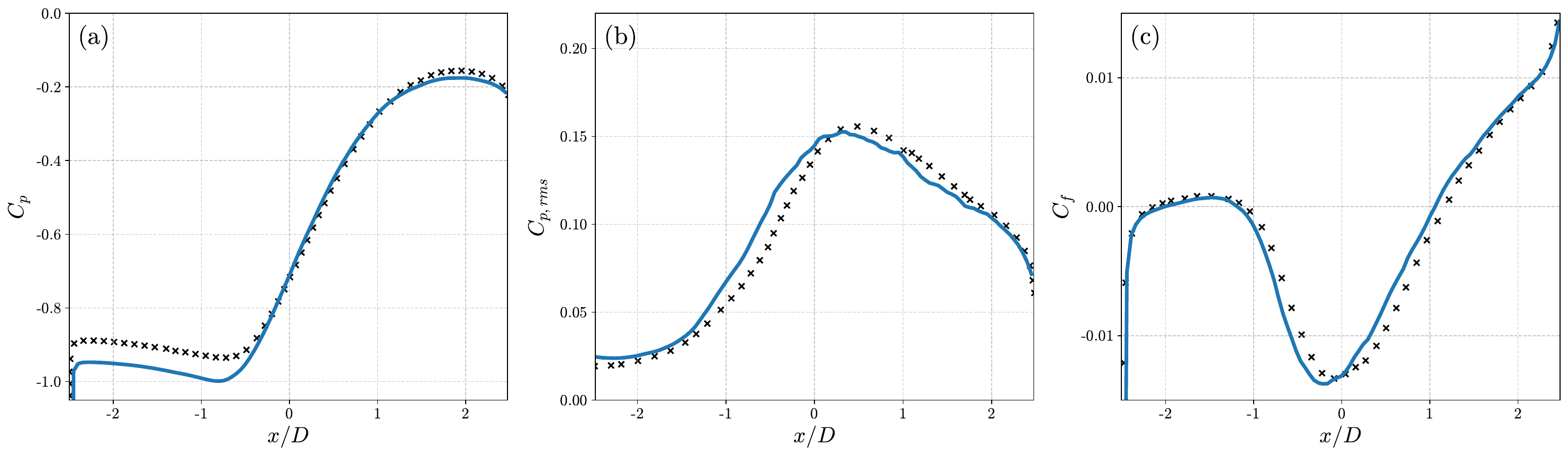}
    \caption{Comparison of (a) mean pressure coefficient distribution, (b) root-mean-square pressure coefficient fluctuations, and (c) skin-friction coefficient over the rectangle surface. Validation data are taken from \citet{cimarelli_direct_2018}.}
    \label{fig:rectsurf}
\end{figure}

\begin{figure}[t]
    \centering
    \begin{minipage}[b]{0.48\textwidth}
        \centering
        \includegraphics[width=\textwidth]{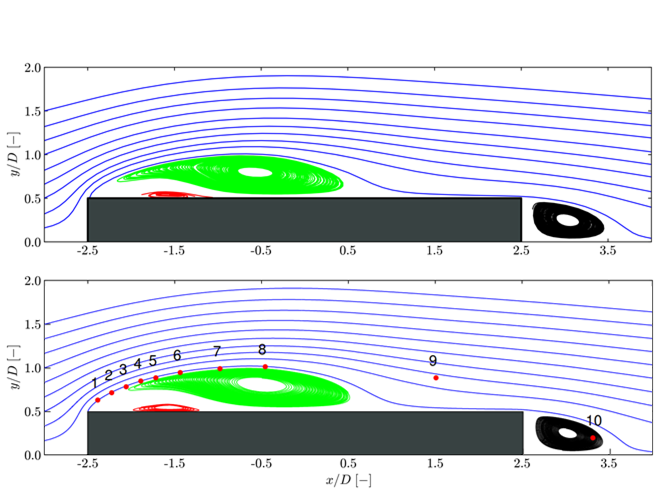}
        \\ (a) 
    \end{minipage}
    \hfill
    \begin{minipage}[b]{0.48\textwidth}
        \centering
        \includegraphics[width=\textwidth]{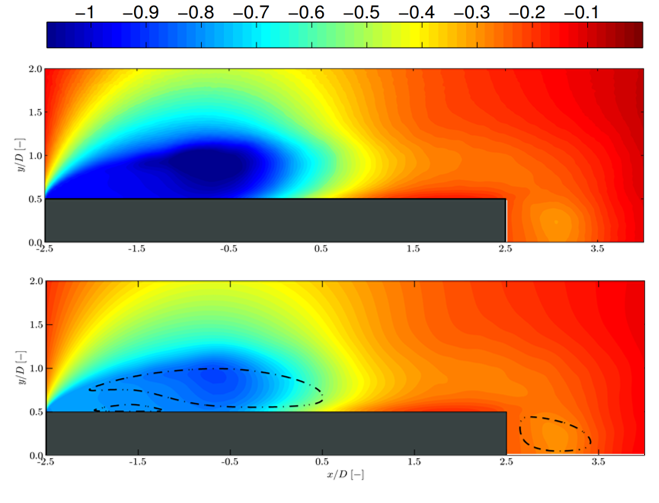}
        \\ (b) 
    \end{minipage}
    \caption{Comparison between the present study (top row) and \citet{cimarelli_direct_2018} (bottom row) of the mean flow field above the rectangular surface. Streamlines of the mean velocity field are shown in (a), and the mean pressure field in (b). In (a), green lines indicate the \textit{primary vortex}, red lines the \textit{secondary vortex}, and black lines the \textit{wake vortex}.}
    \label{fig:cimcom}
\end{figure}

The simulation setup and selected meshing parameters are further evaluated using the Benchmark on the Aerodynamics of a Rectangular 5:1 Cylinder (BARC). The reference data for this case are taken from \citet{cimarelli_direct_2018}, who performed DNS for this configuration. Although this specific rectangle is not included in the dataset, where elongations are restricted to $a \le 1.5$, it provides a strict test case on the method's ability to handle object elongations. In contrast to the circular cylinder, the sharp corners induce fixed separation points while the extended flat top and bottom surfaces offer insight into the simulation's ability to capture near-wall dynamics. To this end, the mean and root-mean-square distributions of the surface pressure coefficient are presented in \autoref{fig:rectsurf}, together with the skin friction coefficient, defined as $C_f = \tau_w / (\frac{1}{2} \rho u_{\infty}^2)$, where $\tau_w = \mu (\partial u / \partial y)_{\text{wall}}$. Overall, the profiles closely match with the DNS data, except for an under-prediction of the mean pressure over the first half of the rectangle. The mean pressure field in \subfigref{fig:cimcom}{b} and streamlines of the time-averaged velocity in \subfigref{fig:cimcom}{a} indicate this discrepancy is located in the region near the \textit{primary vortex} and \textit{secondary vortex}, for which the present simulations predict a more intense region of negative pressure. All other key phenomena, including the \textit{wake vortex}, are accurately captured in both position and magnitude of the associated pressure field.

\subsection*{Square rectangle ($a = 0.5$, $n = \infty$) at $Re = 2000$}
\begin{figure}[t]
    \centering
    \includegraphics[width=\linewidth]{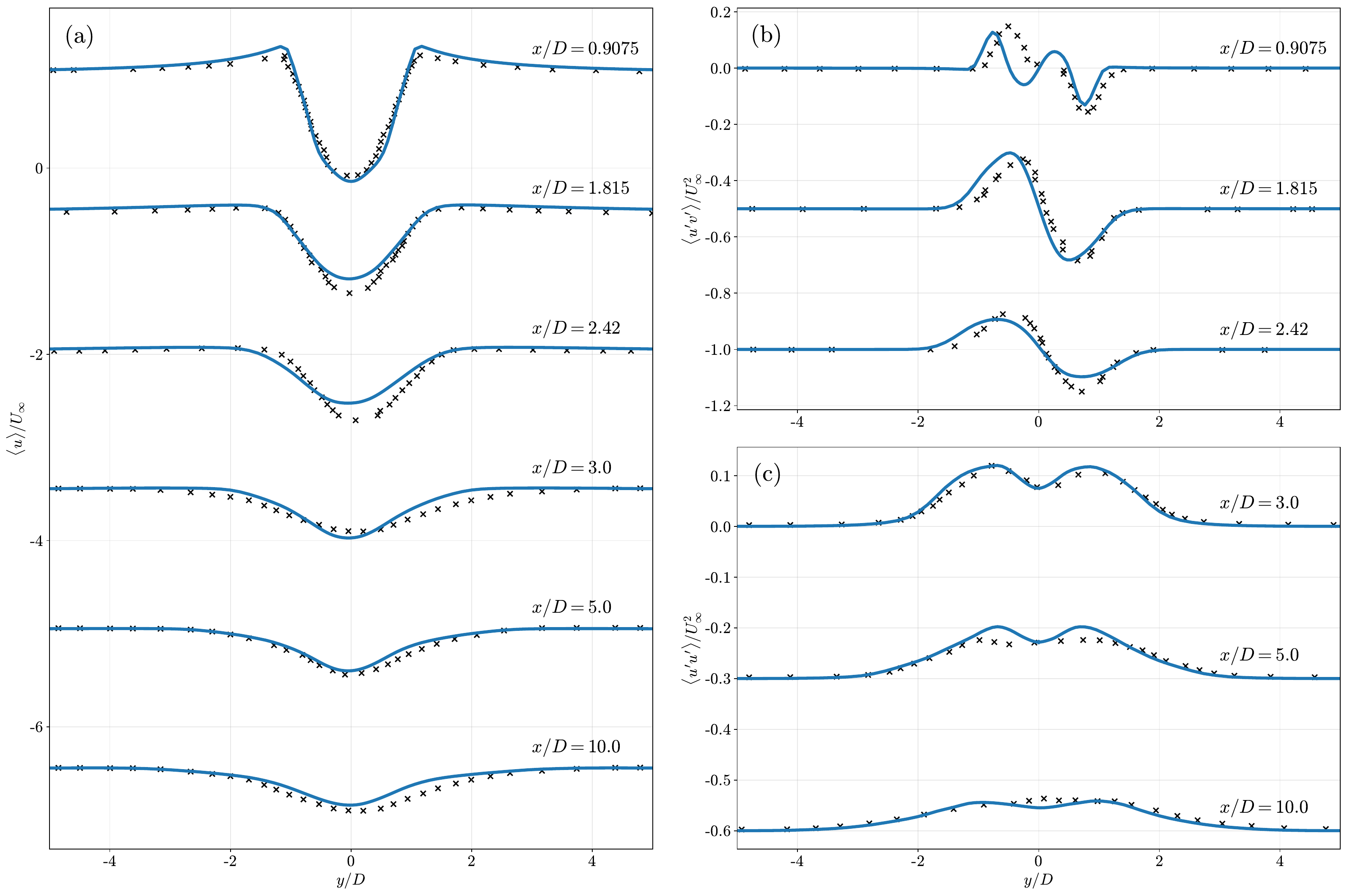}
    \caption{Mean velocity profiles in (a) from near to far downstream of the object, and Reynolds stress components: (b) $u'v'$ close to the cylinder and (c) $u'u'$ in the far wake region. Validation data are taken from \citet{sharma_flow_2022}.}
    \label{fig:squarewake}
\end{figure}

To further assess the performance of the numerical setup in the wake of sharp-edged geometries, simulations are conducted for the square cylinder at $Re = 2000$ and validated against experimental data from \citet{sharma_flow_2022}. \autoref{fig:squarewake} provides a comparison of mean velocity and Reynolds stress profiles at various locations in the wake. Overall, the mean velocity deficit is accurately captured in the near wake and further downstream. A comparable agreement is observed for the Reynolds stresses close to the square. Notably, the simulation predicts a four-peak structure for $\langle u'v'\rangle$ at $x/D = 0.9075$. A comparable result is reported by \citet{sharma_flow_2022} for a confined region around $x/D \approx 0.85$, beyond which the Reynolds stresses transition to the two-peak structure observed in \subfigref{fig:squarewake}{b}. The present results therefore reproduce the four-peak behavior, but its location is slightly shifted relative to the experiments. Finally, Reynolds stress profiles far downstream are accurately reproduced. 

\section{Data Processing} \label{sec:appproc}
After initialization, the simulation contains an unphysical numerical transient. To ensure the solution trajectories are unaffected by this startup phase, a numerical convergence criterion is applied following \citet{Boudet2015}. Measurement probes are placed in regions where the flow exhibits low frequency dynamics. In the present study, they are located along the centerline ($y/D = 0$) at streamwise distances \textcolor{revcolor}{$4D$, $6.5D$ and $9.5D$} behind the object, denoted by $P_{4}$, $P_{6.5}$ and $P_{9.5}$, respectively. These positions correspond to the wake region, which provides a consistent reference across all flow cases, whereas other regions of interest, such as the shear layer, change with geometry. Each probe measures a time-dependent signal, $\phi$. At time $t$, the signal is cut into four equal segments. The final three segments are then compared using the following criterion:
\begin{equation}
    \chi(\phi) = \max \left[ \max_{i,j=2,4} \left(\left| \frac{\overline{\phi}^{(i)}-\overline{\phi}^{(j)}}{\overline{{\phi}}^{(4)}}\right|  \right), \max_{i,j=2,4} \left(\left| \frac{{\phi'}^{(i)}-{\phi'}^{(j)}}{{\phi'}^{(4)}} \right|  \right)    \right].
\end{equation}
The superscripts indicate the temporal segment the data originates from, while $\overline{\phi}$ and $\phi'$ represent the spanwise-averaged mean and standard deviation of the signal, respectively. Once $\chi$ settles around a certain percentage, the flow’s statistics have stabilized over the three most recent temporal quarters, implying the numerical transient was evacuated after the first quarter. Nevertheless, data sampling starts only after the fourth quarter to provide an additional margin of confidence. 

\begin{figure}[t]
    \centering
    \includegraphics[width=\linewidth]{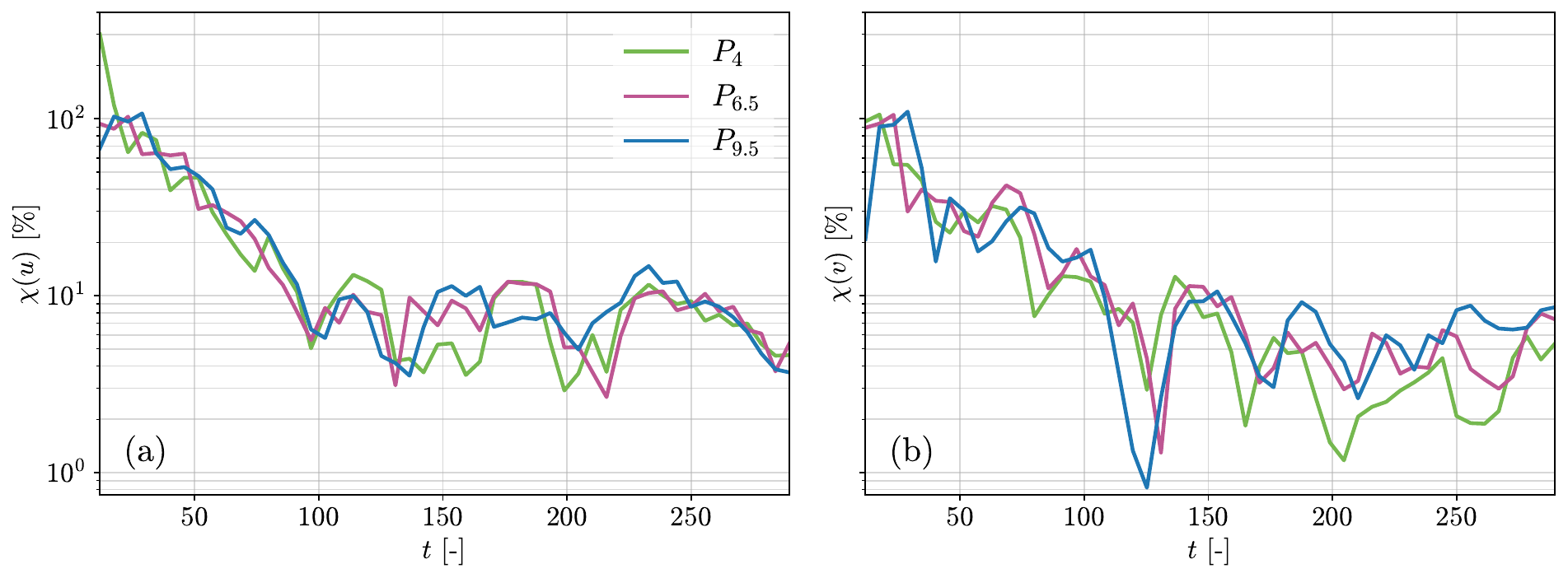} %
    \caption{Numerical convergence criterion $\chi$ for (a) streamwise and (b) vertical velocity at probes $P_{4}$, $P_{6.5}$, and $P_{9.5}$ for the $[Re, \alpha, a, n] = [1796, 11, 0.424, 1]$ case. For $v$, $\chi$ is computed using standard deviation only. Stabilization indicates the evacuation of numerical transients and the onset of statistically steady flow.}
    \label{fig:convcrit}
\end{figure}

The convergence criterion method is initially evaluated after four flow-through times (FTTs) to ensure enough samples are collected for a reliable assessment.  If the signal variations have not yet reached a steady level, the simulation is extended for one FTT at a time until they have. For illustration, \autoref{fig:convcrit} displays the metric for the $u$ and $v$ velocity components for the $[Re, \alpha, a, n] = [1796, 11, 0.424, 1]$ case. Note that for the vertical velocity, only the standard deviation is considered, as its mean approaches zero, which would cause the criterion to diverge. 

Once the numerical transient has been evacuated, the solution fields are sampled at a non-dimensional time-step of $\Delta t_s = 0.04732$. While the solver requires much smaller temporal steps to maintain numerical stability and accuracy, saving the full resolution is computationally prohibitive due to storage requirements. Sub-sampling the simulation at a coarser time-step inevitably results in a loss of information. To quantify the extent of this loss, the energy spectrum is examined. \autoref{fig:cylen071} and \autoref{fig:cylen300} display the energy spectra for the circular cylinder flow case at $Re = 3900$, evaluated at $(x/D, y/D) = [(0.71, 0.66), (3.0, 0.0)] $, computed using the original $u$-velocity signal and the sub-sampled solution at $\Delta t_s = 0.04732$. The spectra match closely across most of the frequencies. In particular, all major characteristic frequencies are accurately recovered. Beyond the inertial subrange, however, the sub-sampled signal is unable to represent the highest frequencies due to its coarser temporal resolution. This is deemed acceptable as this portion of the spectrum is primarily influenced by numerical dissipation rather than physical processes and contains relatively low energy. \textcolor{revcolor}{Note that the sub-sampling acts in time only: each stored snapshot retains the full spatial resolution, so the small-scale spatial content scored by the spectral metric of \autoref{sec:results} is preserved in full.}

{\color{revcolor}
\paragraph{Per-case wake-probe spectra.}
In addition to the convergence probes $P_{4}$, $P_{6.5}$ and $P_{9.5}$ described above, three further centerline probes, denoted $P'_{2.5}$, $P'_{3.5}$ and $P'_{4.5}$, are placed $2.5D$, $3.5D$ and $4.5D$ behind the trailing edge of each object ($y/D=0$). The probes' absolute location thus varies with $a$, $n$, and $\alpha$ across cases. A velocity power spectrum of $u$ and $v$ is computed at each probe for every one of the 400 cases from the 400 saved snapshots of the sub-sampled trajectory ($\Delta t_s$). The resulting spectra are released with each case in the public dataset, together with the script used to compute them. They provide per-case, physically grounded evidence of the turbulent energy cascade as well as an evaluation target for surrogate models beyond the aggregate metrics of \autoref{sec:results}.
}

\begin{figure}[t]
    \centering
    \color{revcolor}
    \includegraphics[width=\linewidth]{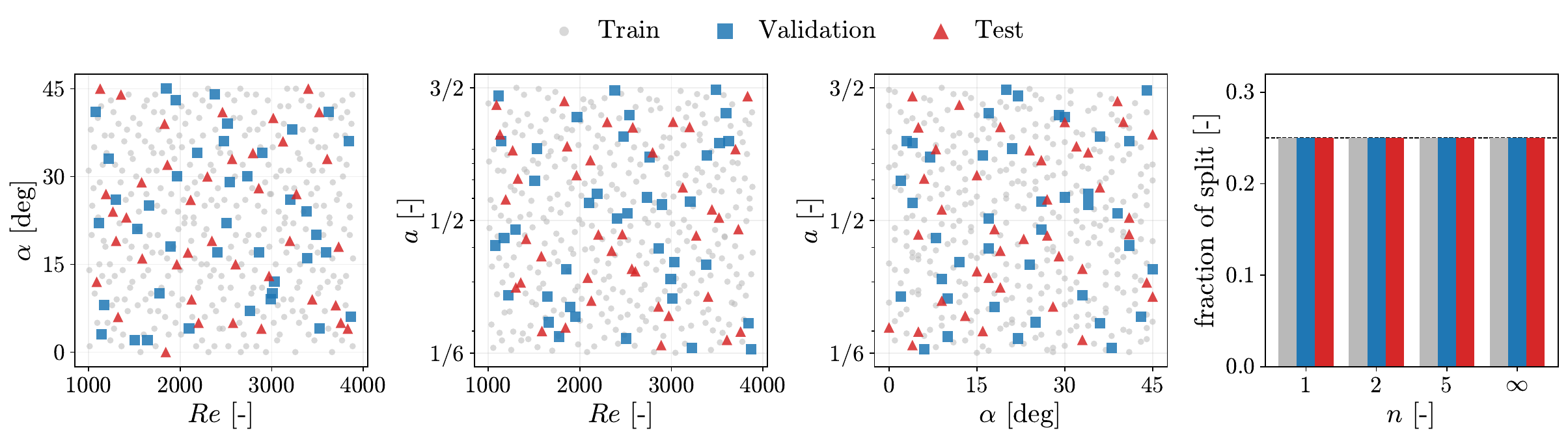}
    \caption{Coverage of the parameter space by the train/validation/test split. The first three panels show the sampled cases projected onto the $(Re, \alpha)$, $(Re, a)$ and $(\alpha, a)$ planes, with training cases in gray, validation cases as blue squares and test cases as red triangles. The right-most panel shows the fraction of each subset belonging to each shape family $n$.}
    \label{fig:splitcoverage}
\end{figure}

{\color{revcolor} \section{Dataset splits} \label{sec:appsplit}
\textsc{TurtLES} is split at the trajectory level into 320 training, 40 validation and 40 test trajectories (80/10/10) such that no trajectory is shared between subsets. Each shape family $n \in \{1, 2, 5, \infty\}$ contains 100 trajectories and is split exactly 80/10/10. Within each family, the assignment of trajectories to subsets is optimized so that the distribution of $(Re, \alpha, \log a)$ in every subset matches that of the full dataset. This prevents a held-out set from concentrating in a corner of the parameter space, which at these subset sizes (40 trajectories) is otherwise likely. The circular-cylinder reference case ($Re = 1000$, $a = 0.5$, $\alpha = 0^\circ$, $n = 2$) is distributed separately and is not part of the split. The result is an in-distribution interpolation split. Extrapolation across geometry, Reynolds number or angle of attack is not assessed here, but is straightforward to construct from the released data: for example, by holding out an entire shape family, or all cases above a Reynolds-number threshold.

\autoref{tab:split_composition} reports the size and geometric composition of each subset, and \autoref{tab:split_parameters} its coverage of the continuous parameters. The shape-family proportions of every subset are identical to those of the full dataset (25\% each), and the statistics of $Re$, $\alpha$ and $a$ are likewise reproduced in each subset. \autoref{fig:splitcoverage} visualizes the sampled parameter space color-coded by split.
}

\begin{table}[t]
    \centering
    \color{revcolor}
    \caption{Size and geometric composition of the \textsc{TurtLES} train/validation/test split. Each subset reproduces the composition of the full dataset.}
    \label{tab:split_composition}
    \small
    \setlength{\tabcolsep}{5.5pt}
    \begin{tabular*}{\linewidth}{@{\extracolsep{\fill}}lcccccc@{}}
        \toprule
        & \multicolumn{2}{c}{Size}
        & \multicolumn{4}{c}{Shape family (\% of split)} \\
        \cmidrule(lr){2-3}
        \cmidrule(l){4-7}
        Split
        & Trajectories
        & Fraction (\%)
        & $n = 1$
        & $n = 2$
        & $n = 5$
        & $n = \infty$ \\
        \midrule
        Train & 320 & 80.0 & 25.0 & 25.0 & 25.0 & 25.0 \\
        Validation & 40 & 10.0 & 25.0 & 25.0 & 25.0 & 25.0 \\
        Test & 40 & 10.0 & 25.0 & 25.0 & 25.0 & 25.0 \\
        \midrule
        All & 400 & 100.0 & 25.0 & 25.0 & 25.0 & 25.0 \\
        \bottomrule
    \end{tabular*}
\end{table}

\begin{table}[t]
    \centering
    \color{revcolor}
    \caption{Parameter coverage of each split. Entries report the mean $\pm$ standard deviation and the $[\min, \max]$ range of the Reynolds number $Re$, the angle of attack $\alpha$, and the super-ellipse semi-axis parameter $a$.}
    \label{tab:split_parameters}
    \small
    \setlength{\tabcolsep}{4pt}
    \begin{tabular*}{\linewidth}{@{\extracolsep{\fill}}lcccccc@{}}
        \toprule
        & \multicolumn{2}{c}{$Re$ [-]}
        & \multicolumn{2}{c}{$\alpha$ [deg]}
        & \multicolumn{2}{c}{$a$ [-]} \\
        \cmidrule(lr){2-3}
        \cmidrule(lr){4-5}
        \cmidrule(l){6-7}
        Split
        & mean $\pm$ std & range
        & mean $\pm$ std & range
        & mean $\pm$ std & range \\
        \midrule
        Train & 2456 $\pm$ 839 & [1001, 3889] & 22.6 $\pm$ 13.2 & [0, 45] & 0.603 $\pm$ 0.365 & [0.167, 1.499] \\
        Validation & 2434 $\pm$ 847 & [1077, 3868] & 22.5 $\pm$ 13.3 & [2, 45] & 0.614 $\pm$ 0.393 & [0.172, 1.476] \\
        Test & 2431 $\pm$ 859 & [1089, 3832] & 22.4 $\pm$ 13.3 & [0, 45] & 0.604 $\pm$ 0.368 & [0.178, 1.399] \\
        \midrule
        All & 2451 $\pm$ 839 & [1001, 3889] & 22.5 $\pm$ 13.2 & [0, 45] & 0.604 $\pm$ 0.367 & [0.167, 1.499] \\
        \bottomrule
    \end{tabular*}
\end{table}

{\color{revcolor}
\section{\textsc{mini-TurtLES}: standardized reduced-size subsets} \label{sec:appmini}

The full dataset is 14~TB, which is prohibitive for many research groups. We therefore define \textsc{mini-TurtLES}, a collection of standardized subsets that reduce the dataset along two independent axes: geometry and time. Fixing the subset definitions ensures that experiments performed in different laboratories use identical data, facilitating reproducibility and fair comparison across published results.

Some storage floor is unavoidable: resolving three-dimensional turbulence requires dense meshes and a sampling interval fine enough to retain the inertial subrange, and reducing either removes the physical content that distinguishes \textsc{TurtLES} from existing benchmarks.

\paragraph{Geometry tiers (G4--G1).}
These tiers reduce the number of obstacle geometry families while retaining the full range of Reynolds numbers ($Re$), angles of attack ($\alpha$), and the geometric semi-axis parameter ($a$). G4 contains all four families $n \in \{1, 2, 5, \infty\}$; G3 removes the $n = 5$ family; G2 further removes $n = 1$; and G1 retains only the elliptical family ($n = 2$). Each tier therefore covers the same physical parameter space with progressively fewer geometries. The ordering follows increasing simplicity while preserving representative flow behavior: $n = 5$ is removed first because it exhibits the least distinctive dynamics, whereas $n = 2$ is retained in every tier because it is the best-studied and most widely used benchmark geometry.

\paragraph{Temporal tiers (T100--T400).}
These tiers retain only the first 100, 200, 300, or 400 snapshots of each trajectory. The geometry and temporal tiers are independent and can be combined freely (e.g.\ G2-T200 or G1-T400). \autoref{tab:minitiers} reports the approximate download size of each subset.

\begin{table}[t]
  \centering
  \color{revcolor}
  \caption{Approximate download size of each \textsc{mini-TurtLES} subset. Geometry tiers
  reduce the number of shape families; temporal tiers truncate each trajectory.}
  \label{tab:minitiers}
  \small
  \begin{tabular}{llcccc}
    \toprule
    Geometry tier & Families $n$ & T100 & T200 & T300 & T400 \\
    \midrule
    G4 & $\{1, 2, 5, \infty\}$ & 3.5 TB & 7.0 TB & 10.5 TB & 14 TB \\
    G3 & $\{1, 2, \infty\}$    & 2.6 TB & 5.1 TB & 7.7 TB & 10.2 TB \\
    G2 & $\{2, \infty\}$       & 1.8 TB & 3.6 TB & 5.3 TB & 7.1 TB \\
    G1 & $\{2\}$               & 0.8 TB & 1.6 TB & 2.4 TB & 3.3 TB \\
    \bottomrule
  \end{tabular}
\end{table}

\paragraph{Debug subset.}
We further provide \textsc{TurtLES-debug} (approximately 5~GB), containing the first 100 snapshots of the circular-cylinder reference case ($Re = 1000$, $a = 0.5$, $\alpha = 0^\circ$, $n = 2$), which is distributed separately from the 400-case split. It is intended for installation checks and dataloader validation.

\paragraph{Retrieval.}
All subsets have fixed names and are described on the dataset page, which details each tier and explains the procedure for downloading it.}


\raggedbottom
{\color{revcolor}
\section{Data visualizations} \label{app:datavis}
To visualize the flows in \textsc{TurtLES}, \autoref{fig:appviews} shows the instantaneous vortical structures of eight cases. For each shape family, one case is shown at low ($Re \leq 1300$) and one at high Reynolds number ($Re > 3000$). Note that at low $Re$ the wake consists of fewer, thicker and more coherent vortex tubes, whereas at high $Re$ it breaks up into a much denser field of fine-scale structures.

\begin{figure}[H]
    \centering
    \color{revcolor}
    \includegraphics[width=\linewidth]{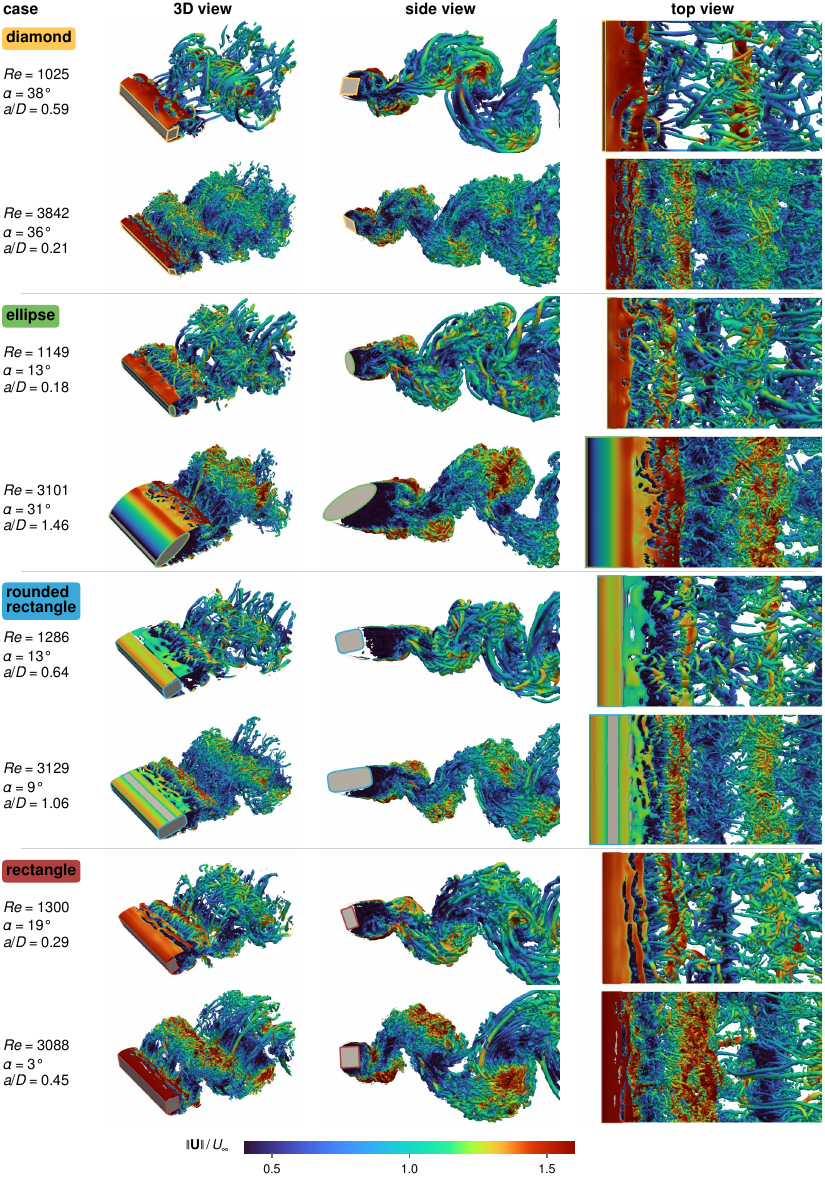} %
    \caption{\textcolor{revcolor}{Visualization of eight \textsc{TurtLES} cases. For each case, 3D, side and top views show isosurfaces of the Q-criterion \citep{hunt1988eddies} ($Q^* = QD^2/U_\infty^2 = 1$) at an instantaneous snapshot, colored by the non-dimensional velocity magnitude $\|\mathbf{U}\|/U_\infty$.}}
    \label{fig:appviews}
\end{figure}
}


\section{Baseline implementations} \label{app:baselines}             
This section documents the architectures, hyperparameters, and training recipes used for the four neural-operator baselines reported in \autoref{sec:results}. We first describe the components shared across baselines, then provide a per-model architectural summary. A consolidated view of the hyperparameters appears in \autoref{tab:baseline_hparams}.                                                                          
\paragraph{Shared training and inference setup.} All baselines are trained on the same temporal task: predict the residual $\Delta u_{t+1}=u_{t+1}-u_t$ from a single context frame $u_t$, conditioned on the Reynolds number~$Re$. Training uses a standard mean-squared-error loss on the residual. We sample one window per gradient step from a uniform-random subset of $524{,}288$ mesh nodes per trajectory, repeated for $100{,}000$ gradient steps with an effective batch size of~$2$ on a single H100 GPU (80~GB). \textcolor{revcolor}{All four baselines are trained on the 320 training trajectories and evaluated on the 40 test trajectories of the split defined in \autoref{sec:appsplit}.} The optimizer is AdamW with learning rate $10^{-4}$, minimum learning rate $10^{-7}$, weight decay $10^{-3}$, $\beta_1=0.9$, $\beta_2=0.95$, and gradient-norm clipping at $1.0$. The schedule is a linear warmup over the first $5{,}000$ steps followed by a cosine decay. PDE-parameter conditioning uses only~$Re$; geometry parameters $(\alpha,a,n)$ are not exposed to the model and are absorbed into the spatial coordinates and field statistics. For all models except GINO we enable bfloat16 mixed precision and \texttt{torch.compile}. GINO's FNO backbone relies on complex-valued spectral convolutions that interact poorly with bfloat16 autocast and produce graph-breaking warnings under \texttt{torch.compile} via the \texttt{neuraloperator}/\texttt{tltorch} factorized-tensor indexing path. We therefore train GINO in fp32 without \texttt{torch.compile}, with a per-GPU batch size of one and gradient accumulation of two to keep the effective batch size at two. Activation checkpointing is enabled for all transformer backbones. \textcolor{revcolor}{Apart from the precision and gradient-accumulation adjustments required by GINO, all four baselines follow an identical protocol: the same residual objective, the same node subsample per gradient step, and the same optimizer, schedule, step count, effective batch size and single-GPU budget. No baseline received additional hyperparameter tuning.} At inference time we use a batch size of one and a context window of one frame. The full mesh (\textcolor{revcolor}{$2.9$--$10.5$}~million nodes per trajectory) is processed either by the model's encoder--decoder split (UPT, GINO, CALM-PDE) or by tile partition (Transolver++) with a chunk size of $524{,}288$ output points. Each test trajectory is unrolled for $399$ autoregressive steps starting from the first ground-truth frame.    
  
\paragraph{UPT~\citep{Alkin2024}.} UPT is a perceiver-style transformer for unstructured meshes. The mesh is first pooled to $K=16{,}384$ supernodes via a runtime $k$-nearest-neighbor gather ($k=32$) and a small message-passing MLP. Two encoder transformer blocks then operate on the supernode tokens. A perceiver cross-attention compresses the supernodes to $L=4{,}096$ learned latent tokens, on top of which we run four self-attention blocks. Two more transformer blocks act on the decoder side before a perceiver cross-attention queries the original mesh coordinates with sincos positional embeddings (max angular frequency $100$) and produces the per-node residual. The hidden dimension is $384$ with $6$ attention heads, an MLP expansion ratio of $4$, no dropout, and DiT-style adaLN-zero conditioning on $Re$. The trained model has $61.96$~M parameters.                                                          

\paragraph{GINO~\citep{li2023geometry}.} GINO consists of a graph-neural-operator (GNO) encoder that maps the irregular mesh onto a regular latent grid, an FNO operating on that grid, and a GNO decoder that queries the predicted residual at the original mesh nodes. The latent grid has resolution~$12^3$ in $[0,1]^3$, with input coordinates min--max normalized to that cube. Both GNOs use a linear transform with neighborhood radius $0.033$ (about one cell at this resolution). The FNO has four spectral layers, $6{\times}6{\times}6$ Fourier modes, $64$ hidden channels, and lifting/projection MLPs with channel ratios of $2$ and $4$ respectively. PDE-parameter conditioning is realized through an AdaIN layer ($4$ conditioning features). We use the official \texttt{neuraloperator} implementation with Open3D-accelerated GNO kernels \citep{kossaifi2025librarylearningneuraloperators}. The trained model has $3.10$~M parameters. \textcolor{revcolor}{This size reflects the shared training budget rather than a tuned choice. GINO is by some margin the slowest baseline per gradient step, because the complex-valued spectral convolutions of its FNO backbone preclude bfloat16 autocast and \texttt{torch.compile} and therefore run in fp32 with gradient accumulation. Within the shared budget of $100{,}000$ steps ($\sim$20~h on a single H100) the model had not converged even at $3.10$~M parameters, and larger configurations were less accurate for the same wall-clock time. The reported configuration is therefore the best achievable under the shared budget.}
                                                         
\paragraph{Transolver++~\citep{luotransolverplus}.} Transolver++ is a monolithic transformer that aggregates $N$ spatial tokens into a small set of learned physics-aware slices via Gumbel-softmax routing, applies self-attention on the slice tokens, then scatters back to the mesh. We use $8$ transformer blocks with hidden dimension $512$, $8$ attention heads, MLP ratio $4$, $128$ slices per head, GELU activation, and no dropout. PDE-parameter conditioning is an additive bias on the encoder output. \textcolor{revcolor}{Because Transolver++ is monolithic, with input and output cardinality equal by construction, partitioning the output necessarily partitions the input as well; there is no latent bottleneck at which the two can be split. Its inference path therefore uses a tile partition rather than an encoder--decoder split: disjoint chunks of the mesh are processed independently, which bounds memory but removes cross-chunk context. This is a consequence of the architecture at this mesh size.} The trained model has $21.79$~M parameters. \textcolor{revcolor}{Transolver++ was developed for steady-state prediction rather than autoregressive dynamics. We include it because it represents the current state of the art for PDE surrogates on million-point geometries, which is one of the defining constraints of this dataset.}

\paragraph{CALM-PDE~\citep{calmpde,koupai2025efficientgenerativetransformeroperators}.} We adopt the continuous-convolution encoder/decoder of CALM-PDE combined with the scaling improvements introduced in ECHO; the implementation is registered as \texttt{CALM-PDE} in our codebase. The encoder uses a continuous convolution with $k{=}32$ neighbors and a $64$-wide kernel MLP to lift features from the irregular mesh to a fixed $16^3 = 4{,}096$-point regular latent grid in $[0,1]^3$, with global coordinate bounds computed once over the \textcolor{revcolor}{320 training trajectories} so that the latent grid is shared across batches. A UPT-style transformer with hidden dimension $384$, $6$ heads, MLP ratio $4$, and $8$ blocks acts as the time-stepping operator on the latent tokens, conditioned on $Re$ via DiT adaLN-zero. The continuous convolution decoder maps the post-transformer latent back to the original mesh nodes, with output queries chunked at $32{,}768$ points to bound peak memory. The trained model has $45.14$~M parameters.

\begin{table}[t]                           
      \centering                                         
      \caption{Summary of baseline hyperparameters. Counts are per model.}   
      \label{tab:baseline_hparams}
      \small                      
      \setlength{\tabcolsep}{4pt}                        
      \begin{tabular*}{\linewidth}{@{\extracolsep{\fill}}lcccc@{}}   
          \toprule 
          Hyperparameter & UPT & GINO & Transolver++ & CALM-PDE \\   
          \midrule      
          Trainable parameters & $61.96$~M & $3.10$~M & $21.79$~M & $45.14$~M \\    
          Hidden dim & $384$  & $64$ (FNO) & $512$  & $384$ \\   
          Heads & $6$    & -- & $8$    & $6$   \\   
          MLP ratio & $4$    & -- & $4$    & $4$   \\    
          Depth & 2/4/2 (enc/lat/dec) & $4$ (FNO) & $8$ & $8$ \\     
          Tokens / latent grid & $K{=}16{,}384$, $L{=}4{,}096$ & $12^3$ & $128$ slices & $16^3{=}4{,}096$ \\         
          $k$-NN $k$ & $32$   & -- & --     & $32$  \\       
          Conditioning & adaLN-zero ($Re$) & AdaIN ($Re$) & additive bias ($Re$) & adaLN-zero ($Re$) \\ 
          Activation checkpointing & yes    & --  & yes    & yes \\ 
          Mixed precision (bf16) & yes & no   & yes    & yes   \\   
          \texttt{torch.compile} & yes & no & yes    & no    \\      
          Steps & \multicolumn{4}{c}{$100{,}000$} \\     
          Optimizer & \multicolumn{4}{c}{AdamW ($\beta_1{=}0.9$, $\beta_2{=}0.95$, wd${=}10^{-3}$)} \\ 
          Learning rate & \multicolumn{4}{c}{$10^{-4}$ $\rightarrow$ $10^{-7}$, cosine, $5{,}000$ linear warmup} \\            
          Effective batch size & \multicolumn{4}{c}{$2$ (GINO: bs${=}1$, accum${=}2$)} \\ 
          Subsampled nodes / step&\multicolumn{4}{c}{$524{,}288$} \\ 
          Hardware & \multicolumn{4}{c}{$1{\times}$ NVIDIA H100 80~GB, ${\sim}20$~h wall time} \\ 
          \bottomrule  
      \end{tabular*} 
  \end{table}

\newpage

\end{document}